\documentclass[aps,prd,superscriptaddress,twocolumn,showpacs,eqsecnum,nofootinbib,floatfix]{revtex4-1}

\pdfoutput=1
\usepackage{amssymb}
\usepackage{gensymb}
\usepackage{graphicx}
\usepackage{amsmath}
\usepackage{wasysym}
\usepackage{verbatim}

\begin{document}
\title{The Sensitivity of HAWC to High-Mass Dark Matter Annihilations}

\author{A. U. Abeysekara}
\affiliation{Department of Physics and Astronomy, Michigan State University, East Lansing, MI, USA}
\author{R. Alfaro}
\affiliation{Instituto de F{\'\i}sica, Universidad Nacional Aut{\'o}noma de M{\'e}xico, Mexico D.F., Mexico}
\affiliation{Department of Physics, University of Maryland, College Park, MD, USA}
\author{C. Alvarez}
\affiliation{CEFyMAP, Universidad Aut{\'o}noma de Chiapas, Tuxtla Guti{\'e}rrez, Chiapas, M{\'e}xico}
\affiliation{Department of Physics, University of Maryland, College Park, MD, USA}
\author{J. D. {\'A}lvarez}
\affiliation{Universidad Michoacana de San Nicol{\'a}s de Hidalgo, Morelia, Mexico}
\author{R. Arceo}
\affiliation{CEFyMAP, Universidad Aut{\'o}noma de Chiapas, Tuxtla Guti{\'e}rrez, Chiapas, M{\'e}xico}
\author{J. C. Arteaga-Vel{\'a}zquez}
\affiliation{Universidad Michoacana de San Nicol{\'a}s de Hidalgo, Morelia, Mexico}
\author{H. A. Ayala Solares}
\affiliation{Department of Physics, Michigan Technological University, Houghton, MI, USA}
\author{A. S. Barber}
\affiliation{Department of Physics and Astronomy, University of Utah, Salt Lake City, UT, USA}
\author{B. M. Baughman}
\email{bbaugh@umdgrb.umd.edu}
\affiliation{Department of Physics, University of Maryland, College Park, MD, USA}
\author{N. Bautista-Elivar}
\affiliation{Universidad Politecnica de Pachuca, Pachuca, Hgo, Mexico}
% ...

\collaboration{The HAWC Collaboration}
\noaffiliation
\author{K. N. Abazajian}
\email{kevork@uci.edu}
\affiliation{Department of Physics and Astronomy, University of California, Irvine, Irvine, CA, USA}

\date{\today}

\begin{abstract}
The High Altitude Water Cherenkov (HAWC) observatory is a wide field-of-view detector sensitive to gamma rays of 100\,GeV to a few hundred TeV.
Located in central Mexico at $19\degree$ North latitude and 4100\,m above sea level, HAWC will observe gamma rays and cosmic rays with an array of water Cherenkov detectors.
The full HAWC array is scheduled to be operational in Spring 2015.
In this paper, we study the HAWC sensitivity to the gamma-ray signatures of high-mass (multi-TeV) dark matter annihilation.
The HAWC observatory will be sensitive to diverse searches for dark matter annihilation, including annihilation from extended dark matter sources, the diffuse gamma-ray emission from dark matter annihilation, and gamma-ray emission from non-luminous dark matter subhalos.
Here we consider the HAWC sensitivity to a subset of these sources, including dwarf galaxies, the M31 galaxy, the Virgo cluster, and the Galactic center.
We simulate the HAWC response to gamma rays from these sources in several well-motivated dark matter annihilation channels.
If no gamma-ray excess is observed, we show the limits HAWC can place on the dark matter cross-section from these sources.
In particular, in the case of dark matter annihilation into gauge bosons, HAWC will be able to detect a narrow range of dark matter masses to cross-sections below thermal.
HAWC should also be sensitive to non-thermal cross-sections for masses up to nearly 1000 TeV.
The constraints placed by HAWC on the dark matter cross-section from known sources should be competitive with current limits in the mass range where HAWC has similar sensitivity.
HAWC can additionally explore higher dark matter masses than are currently constrained.
\end{abstract}




\maketitle
\section{Introduction}
The effects of dark matter have been seen in many observations: galactic rotation curves, galaxy clusters, gravitational lensing, large-scale cosmological structure, and the cosmic microwave background.
The particle nature of the dark matter remains unclear (for a review of dark matter particle candidates, see, e.g.~\cite{Feng:2010gw}).
Of the candidates which have been considered, the weakly-interacting massive particle (WIMP) is perhaps the best-motivated.
In areas of high dark matter density, WIMPs can annihilate into Standard Model particles and produce photons via pion decay, radiative processes by charged leptons, or direct production of gamma rays through loop-order processes.
The detection of these dark matter annihilation products is referred to as ``indirect detection'' of dark matter, and can be used to constrain the mass, annihilation spectrum, and annihilation cross-section of the dark matter.

To produce the dark matter relic density observed in nature, a thermal relic WIMP should have a weak-scale cross-section of $\langle\sigma_{\rm{A}}v\rangle_0\approx2.2\times10^{-26}\rm\,cm^3\,s^{-1}$ ($\langle\sigma_{\rm{A}}v\rangle_0\approx4.4\times10^{-26}\rm\,cm^3\,s^{-1}$) for a Majorana (Dirac) dark matter particle, largely independent of the dark matter mass~\cite{Steigman:2012nb}.
For comparison to other work, we consider the canonical thermal cross-section $\langle\sigma_{\rm{A}}v\rangle_0\approx3\times10^{-26}\rm\,cm^3\,s^{-1}$ in this paper.
However, some dark matter candidates with multi-TeV masses are not thermally produced but produced through decays of heavier thermally-produced particles or may have their cross-sections enhanced through resonances with heavier, unstable dark matter states~\cite{Feng:2010gw}, so here the thermal cross-section is only a representative benchmark; the dark matter cross-section could be above or below it.

The High Altitude Water Cherenkov (HAWC) observatory is a high-energy gamma-ray observatory currently being installed at Sierra Negra, Mexico.
The site is 4100\,m above sea level, at latitude $18\degree59.7'$N and longitude $97\degree18.6'$W.
The water Cherenkov design has previously been used successfully with the Milagro Gamma-Ray Observatory for observations of the Galactic plane and point sources with emission energies above 1\,TeV~\cite{Abdo:2009ku,Atkins:2004jf}.
HAWC is sensitive to gamma rays of 100\,GeV to a few hundred TeV.
HAWC will consist of a $22,000\rm\,m^2$ array of 300 water tanks.
Each tank will contain four photo-multiplier tubes for observing the Cherenkov light emitted by charged particles passing through the water.

Water Cherenkov detectors work by directly detecting the particles from the extensive air shower associated with a high-energy cosmic ray or gamma ray entering the atmosphere.
At ground level, water Cherenkov detectors measure the Cherenkov light given off inside each detector when charged air shower particles (and gammas converted to $e^+e^-$ in the water) pass through it.
The HAWC design has a wide field of view of 2\,sr, with nearly 8\,sr observable each sidereal day.
The detector can operate continuously, during day and night and regardless of weather.
The effective field-of-view is usually limited to within $45\degree$ of the zenith, but can in principle simultaneously observe photons coming from the entire hemisphere of the sky, with a higher energy threshold for showers coming from larger zenith angles.
This large field-of-view allows the observatory to detect sources in multiple locations at once without pointed observations.

HAWC is expected to be $\sim15$ times more sensitive than Milagro, with a 1-year sensitivity above 2 TeV of $3\times10^{-13}\rm\,cm^{-2}\,s^{-1}$~\cite{Abeysekara:2013tza}.
With its design, HAWC is sensitive to sources extended by several degrees.
The HAWC angular resolution above 100\,TeV is $0.08\degree$, which degrades to $0.8\degree$ below 300\,GeV~\cite{Abeysekara:2013tza}.
Being at high altitude allows the detector to collect significantly more electromagnetic particles in each air shower than at lower altitudes, so HAWC is sensitive to photon energies down to hundreds of GeV.
The logarithmic energy resolution is $\sim100\%$ at low energies, improving to $\sim30\%$ above 30\,TeV~\cite{Abeysekara:2013tza}.

The results presented here employ the detailed simulation of the HAWC detector and extensive air showers.
Air showers produced by gamma rays, protons, helium, and heavier nuclei are simulated using CORSIKA~\cite{citeulike:9245764} and then injected into a detailed detector description within GEANT4~\cite{citeulike:6822207,citeulike:9339116} to simulate the detector response.
The GEANT4 output is reconstructed using the HAWC reconstruction software~\cite{Abeysekara:2013tza}, to characterize the detector sensitivity to particular sources.
In this reconstruction, we do an analysis of the sensitivity of HAWC to dark matter annihilations assuming that further analysis with the detector does not improve our understanding of the HAWC detector response.
However, particularly above $\sim10$ TeV, the HAWC background rejection algorithm is limited by small numbers of Monte Carlo events and is expected to improve once the full detector begins taking data (for more details, see section 2 of Ref.~\cite{Abeysekara:2013tza}).

A search for annihilating dark matter should account for different source classes, which will have different expected dark matter kinematics, dark matter densities, dark matter substructure, and baryonic effects that can affect signal and background levels.
Therefore, a survey of several source populations is necessary to search for annihilating dark matter.
Searches can be conducted on any object that is expected to have high dark matter content, preferably with sources that also have low astrophysical gamma production backgrounds.
In addition, multiple sources can be combined to give stronger evidence for measured dark matter signals.

 The HAWC observatory is sensitive to dark matter annihilation from several source classes.
HAWC has a particular sensitivity to very extended sources (such as galaxies and galaxy clusters) as well as moderately extended sources (such as dwarf galaxies), so HAWC can search the details of dark matter profiles of such extended sources.
The large sky coverage of HAWC is also excellent for the search for diffuse gamma-rays from dark matter annihilation, both from the Galactic halo and from diffuse extragalactic dark matter populations.
With the HAWC sky survey, it is also possible that it could observe gamma-ray emission from nearby dark matter subhalos which have too few stars to be detected optically.

In this work, we perform a forecast of the HAWC sensitivity to signatures of dark matter annihilation in the Coma Berenices dwarf galaxy, the Draco dwarf galaxy, the Segue 1 dwarf galaxy, the M31 galaxy, the Virgo galaxy cluster, and the Galactic center (GC).
The dwarf galaxies provide sources with low gamma-ray backgrounds over a range of declinations and dark matter densities.
The Virgo cluster and M31 are expected to contain appreciable dark matter substructure which should increase the gamma-ray flux from dark matter.
Because substructure tends to dominate further out in sources than the smooth halo, detection of these sources benefits from the HAWC ability to detect extended sources.
The GC has a very low declination for HAWC observations, near the edge of the HAWC field-of-view, and has angular extension on the sky that depends on its dark matter profile.
This is only a small sample of the dark matter sources available to HAWC, but these sources allow us to study the effects of source flux, source declination, dark matter profile, and instrument response on the sensitivity of the detector.
The projected dark matter limits from HAWC are stronger for high-mass WIMPs than most limits coming from lower-energy observatories.

\section{TeV-Scale WIMPs}
The motivation to consider TeV-scale WIMP dark matter candidates has become stronger in recent years.
With null results of dark matter discovery from both the LHC and {\it Fermi}-LAT, it is increasingly likely that the dark matter is comprised of higher-mass WIMPs.
Multi-TeV WIMPs have been considered to explain several recent astrophysical anomalies, which we discuss below.

The recent findings of the AMS-02 experiment, which showed a positron excess rising to over 350\,GeV, have spawned great interest~\cite{Aguilar:2013qda}.
These findings extend the findings of the PAMELA collaboration~\cite{Adriani:2008zr}, which were verified and extended by the {\it Fermi}-LAT collaboration~\cite{FermiLAT:2011ab}.
Annihilation of leptophilic dark matter in a nearby subhalo has been discussed as the possible source of these anomalies~\cite{Cholis:2008wq,Cholis:2008qq,Cirelli:2008jk,Profumo:2009uf,Cholis:2013psa}.
In these models, masses from hundreds of GeV up to nearly 10 TeV are consistent with the local positron excess.
In particular, dark matter models which produce multiple leptons through annihilation into light mediator particles are now favored to produce this excess~\cite{Cholis:2013psa}.

The possibility that the H.E.S.S. observatory observations of the GC show evidence of multi-TeV WIMP annihilation has been considered for some time~\cite{Bergstrom:2004cy,Horns:2004bk,Profumo:2005xd,Cembranos:2013fya}.
In particular, the signal is consistent with a WIMP mass of tens of TeV annihilating primarily into gauge bosons or quarks.
The cross-section for such a WIMP is $\sim1000$ times larger than thermal, due to Sommerfeld enhancement or possibly a non-thermal dark matter.
Both Kaluza-Klein dark matter~\cite{Bergstrom:2004cy} and Branon dark matter~\cite{Cembranos:2013fya} have been considered as possible candidates to produce such a signal.

The discovery of a small-scale ($\sim 10\degree$) anisotropy in the arrival directions of multi-TeV cosmic rays (CRs) has been observed by multiple experiments~\cite{Abdo:2008kr,Aartsen:2012ma,ARGO-YBJ:2013gya}.
Such a signal may require both a non-standard CR propagation in the local neighborhood and an extremely nearby cosmic ray source (within tens of parsecs), both of which would challenge standard assumptions about cosmic-ray sources.
A nearby dark matter subhalo has been discussed as a possible source close to the Earth~\cite{Harding:2013qra}.
It was shown that a subhalo of 20-200 TeV WIMPs dominated by hadronic or bosonic annihilation channels could successfully explain the TeV CR anisotropy if the subhalo were within 100\,pc from the Earth.
Furthermore, the needed WIMP cross-section, mass, and channel for this signal is the same as those which explain the H.E.S.S. GC signal.
In addition to the possibility that HAWC will observe the production of cosmic rays in WIMP annihilations, this hypothesis also implies the existence of a gamma-ray signal which HAWC can observe within the first year of full operation.
The gamma-ray source is not expected to be spatially-coincident with the CR signal, but it should be a very extended source ($\sim 10\degree$ across) with a gamma-ray spectrum consistent with 20-200 TeV WIMP annihilation~\cite{Harding:2013qra}.

In addition to the 20-200 TeV candidates suggested above, PeV-mass dark matter has been considered as well.
Recently, the IceCube observatory has an excess of TeV- and PeV-energy neutrinos above the expected atmospheric background~\cite{Aartsen:2013bka,Aartsen:2013jdh}.
Such high-energy neutrinos could only come from a few source classes, and one that has been suggested is local PeV-scale WIMPs~\cite{Esmaili:2013gha,Bai:2013nga}.
This explanation is consistent with current measurements, and it is not strongly-dependent on the dark matter channel.
Such a WIMP signal should also lead to a relatively large diffuse gamma-ray flux at high energy, which HAWC could measure.

It should be noted that in high-mass WIMP annihilation models above $\sim 100$ TeV, unitarity of the scattering matrix can often give an upper bounds to the dark matter mass~\cite{Griest:1989wd,Hui:2001wy}.
However, some dark matter models do have masses larger than this without violating this bound~\cite{Profumo:2005xd}, so for completeness we consider the HAWC limits up to dark matter masses of 1000 TeV here.

\begin{table*}
\begin{center}
\begin{tabular}[t]{|l|c|c|c|c|c|c|c|}
  \hline
  Parameter & Coma Ber. & Draco & Segue 1 & GC NFW & GC Einasto & M31 (Smooth) & Virgo (smooth)\\
  \hline
  Declination (J2000) & $+23\degree54'15''$ & $+57\degree54'55''$ & $+16\degree04'25''$ & $-29\degree00'28''$ & $-29\degree00'28''$ & $+41\degree16'09''$ & $+12\degree20'13''$\\
  Distance $R$ (kpc) & 44$\rm^a$ & 76$\rm^a$ & 23$\rm^b$ & 8.5$\rm^c$ & 8.28$\rm^c$ & 784$\rm^d$ & 16800$\rm^e$\\
  Scale Density $\rho_s\rm\ \left(\frac{GeV}{cm^{3}}\right)$ & 9.76$\rm^a$ & 0.976$\rm^a$ & 4.18$\rm^b$ & 0.259$\rm^c$ & 0.0715$\rm^c$ & 1.44$\rm^d$ & 0.0189$\rm^e$\\
  Scale Radius $r_s$ (kpc) & 0.16$\rm^a$ & 2.1$\rm^a$ & 0.15$\rm^b$ & 20.0$\rm^c$ & 21.0$\rm^c$ & 8.18$\rm^d$ & 545$\rm^e$\\
 Optimal Bin $\Delta\Omega_{\rm opt}$ (msr) & 0.69$\rm^{f}$ & 0.90$\rm^{f}$ & 0.69$\rm^{f}$ & 8.44$\rm^{f}$ & 61.2$\rm^{f}$ & 0.78$\rm^{f}$ & 1.00$\rm^{f}$ \\
   $J_{\Delta\Omega}(\Delta\Omega_{\rm opt})$ & 1.6 & 7.7 & 10.9 & 444.8 & 201.4 & 12.5 & 0.90\\
  \hline
\end{tabular}
\caption[Declinations and halo parameters for DM sources.]{\label{haloparams}: Declinations and halo parameters for Coma Berenices, Draco, Segue 1, the GC with an NFW profile, the GC with an Einasto profile, the smooth component of M31, and the smooth component of the Virgo cluster.
We also consider substructure-boosted profiles for M31 and the Virgo cluster, with boosts taken from Ref.\cite{Sanchez-Conde:2013yxa}.
For reference, HAWC is located at $18\degree59'41''$ North latitude.
$\rm^a$Coma Berenices and Draco data are chosen as in Ref.~\cite{Abdo:2010ex}.
$\rm^b$Segue 1  data are chosen as in Ref.~\cite{Aliu:2012ga}.
$\rm^c$GC data are chosen as in Ref.~\cite{Abazajian:2010sq}.
$\rm^d$M31 smooth halo parameters are chosen as in Ref.~\cite{Geehan:2005aq}.
$\rm^e$Virgo smooth halo parameters are calculated as in Ref.~\cite{Han:2012uw}.
$\rm^f$Quoted optimal bins are calculated at 1 TeV, at which HAWC has a $0.5\degree$ point-spread function.
}
\end{center}
\end{table*}

\section{Gamma-Ray Emission from Annihilating Dark Matter}
\subsection{Dark Matter Differential Flux}
A calculation of the gamma-ray flux from dark matter annihilation requires information about both the astrophysical properties of the dark matter source and the particle properties of the final-state radiation.
The differential gamma-ray flux integrated over solid angle $\Delta\Omega$ for a dark matter candidate with cross-section (times the relative velocity of the interacting dark matter particles) $\langle\sigma_{\rm{A}}v\rangle_0$ is
\begin{equation}
\frac{dF}{dE}=\frac{\langle\sigma_{\rm{A}}v\rangle_0}{2}\frac{J_{\Delta\Omega}}{J_0}\frac{\Delta\Omega}{4\pi M_\chi^2}\frac{dN_\gamma}{dE}
\end{equation}
where $dN_\gamma/dE$ is the $\gamma$-ray spectrum per dark matter annihilation and $M_\chi$ is the dark matter particle mass.
The mass density ($\rho$) squared integrated along the line-of-sight $x$, averaged over the solid angle of the observation region is defined as
\begin{equation}
J_{\Delta\Omega}=\frac{J_0}{\Delta\Omega}\int_{\Delta\Omega}d\,\Omega\int d\,x\ \rho^2(r_{\rm{gal}}(\theta,x))
\end{equation}
where distance from the source is given by
\begin{equation}
r_{\rm{gal}}(\theta,x)=\sqrt{R^2-2xR\cos(\theta)+x^2}\enspace.
\end{equation}
$J_0\equiv 1/\left[8.5\,{\rm kpc}\ (0.3\,{\rm GeV\,cm^{-3}})^2\right]$ is a normalization constant chosen to make $J_{\Delta\Omega}$ dimensionless, but the final flux calculation is independent of the choice of $J_0$.
$R$ is the distance to the center of the source and $\theta$ is the angle between the line-of-sight and the source.
The specific parameters for the sources we consider here are given in Table~\ref{haloparams}.

\subsection{Effects of Substructure}
With its large field-of-view, HAWC is an ideal detector to search for extended sources of dark matter.
Other galaxies, like M31, and galaxy clusters, including Virgo, have large dark matter halos that should extend out to several degrees in the sky.
HAWC can observe the full dark matter halo for such objects, including the halo regions far from any luminous background.
Additionally, spatially-extended large sources of dark matter are expected to have a dark matter J-factor larger than those in Table~\ref{haloparams}, boosted due to small dark matter substructures in the outlying regions of the halo~\cite{Kamionkowski:2010mi}.
The dark matter substructure dominates far from the center of the dark matter halos, so large field-of-view detectors like HAWC should be sensitive to the increased dark matter flux.

Throughout this work, we assume no annihilation boost factor from dark matter substructure for the dwarf galaxies or the GC.
For M31 and the Virgo cluster, which necessarily have dark matter substructure because they are a galaxy and a cluster of individual galaxies, we consider both the contribution from a substructure-boosted halo as well as from the smooth dark matter halo.
We consider the substructure--boosted model of Ref.~\cite{Sanchez-Conde:2013yxa}, which gives a boost factor of $\sim15$ for M31 and $\sim35$ for the Virgo cluster.
Note that these boost factors are very conservative (for example, the model of Ref.~\cite{Han:2012uw} gives a boost factor of 1000 for the Virgo cluster).
Because the amount of subclustering and its corresponding boost factors are active areas of research, so we include results both with and without subclustering here.
\begin{figure}
\begin{center}
\includegraphics[width=0.5\textwidth]{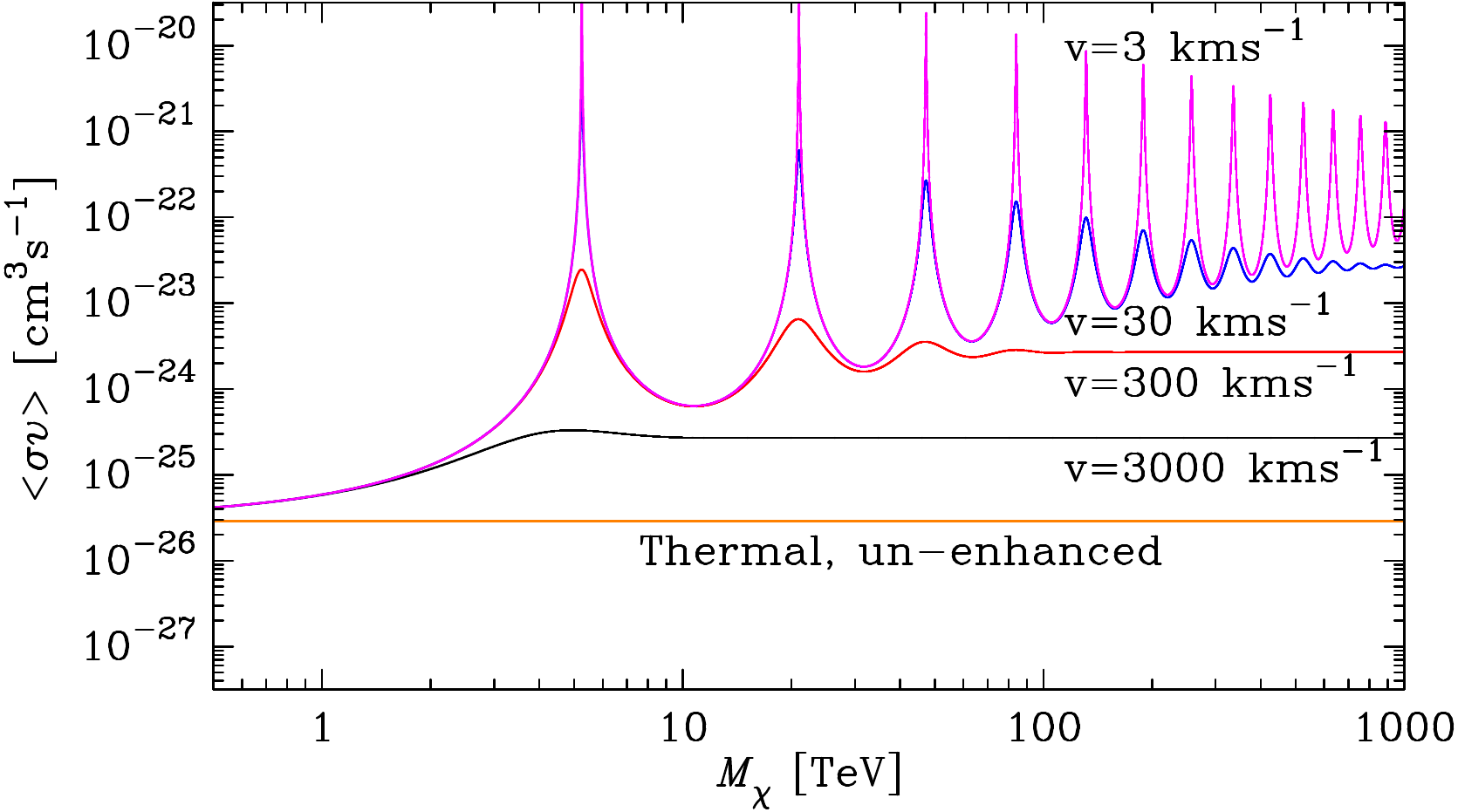}
\end{center}
\caption[Natural Sommerfeld enhancement for several dark matter velocities.]{\small The Sommerfeld enhancement to the dark matter cross-section from the exchange of $W$ bosons.
The calculation uses the formalism of Ref.~\cite{Feng:2010zp} and assumes weak-scale coupling of 1/35.
The dark matter cross-section is shown for 4 dark matter velocities: $3 km\,s^{-1}$ (magenta), $30 km\,s^{-1}$ (blue), $300 km\,s^{-1}$ (red), and $3000 km\,s^{-1}$ (black).
The un-enhanced thermal cross-section ($v=3\times10^5 km\,s^{-1}$) is shown for comparison (purple).
In this work, we assume $v_{\rm rel}=300\rm\,{km\,s^{-1}}$, thought to be the relative velocity of Galactic dark matter.
\label{Sommfig}}
\end{figure}
\subsection{Sommerfeld Enhancement}
For a  thermal relic WIMP, a cross-section of $\langle\sigma_{\rm{A}}v\rangle_0\approx3\times10^{-26}\rm\,cm^3\,s^{-1}$ in the early universe is needed in order to produce the dark matter density observed today.
However, the kinematics of the dark matter today are very different from those in the early universe.
At thermal freeze-out, dark matter annihilated with a relative velocity of $v_{\rm rel}\sim10^{5}\rm\,{km\,s^{-1}}$, whereas Galactic dark matter is thought to interact with $v_{\rm rel}\sim300\rm\,{km\,s^{-1}}$ and dark matter in local clumps could approach $v_{\rm rel}\sim0$.
If the dark matter couples to gauge bosons, this can create a resonance which is amplified for low-velocity dark matter and significantly increases the dark matter cross-section with respect to thermal, a process referred to as Sommerfeld enhancement~\cite{Feng:2010zp}. In this case, a cross-section of $\langle\sigma_{\rm{A}}v\rangle_0$ in the early universe produces a cross-section $\langle\sigma_{\rm{A}}v\rangle$ today which can be much larger.
Typically, the exchanged boson is assumed to exist in the dark sector and be very light; however, for high-mass WIMPs which couple to the $W$ and $Z$ standard-model bosons, there should also be a Sommerfeld enhancement from those interactions~\cite{Lattanzi:2008qa,Aliu:2012ga}.

In this paper, we choose $v_{\rm rel}\sim300\rm\,{km\,s^{-1}}$, which conservatively assumes that the dark matter relative velocity in our sources is identical to the local speed at Earth, though in dwarf galaxies and other dark matter substructures the dark matter speed is expected to be smaller.
As is shown in Figure~\ref{Sommfig}, this choice of dark matter velocity gives a cross-section enhancement of two orders of magnitude above thermal, well below the cross-section enhancements in regions where $v_{\rm rel}\sim0$.
Our choice of coupling between the dark matter and Standard Model gauge bosons is $\alpha_{X}=g^2/4\pi\sim1/35$, which is the Standard Model weak-scale coupling and therefore assumed to be the coupling for weak-scale WIMPs.
We only consider the Sommerfeld enhancement for the $W^+W^-$ dark matter annihilation channel, which is guaranteed to have a dark matter coupling to gauge bosons when the WIMP mass is greater than the mass of the gauge boson~\cite{Lattanzi:2008qa}.
In figures 2-4, we show the limits on the early-universe cross-section $\langle\sigma_{\rm{A}}v\rangle_0$ for the $W^+W^-$ annihilation channel, as well as the early-universe thermal cross-section of $\langle\sigma_{\rm{A}}v\rangle_0\approx3\times10^{-26}\rm\,cm^3\,s^{-1}$ for comparison. For all curves which do not have Sommerfeld enhancement, the cross-sections $\langle\sigma_{\rm{A}}v\rangle_0=\langle\sigma_{\rm{A}}v\rangle$ are identical and used interchangeably.
For calculating the Sommerfeld enhancement as a function of dark matter mass, we use the formalism of Ref.~\cite{Feng:2010zp}.
\subsection{Dark Matter Profiles}
The dark matter profiles $\rho(r)$ used in this paper are chosen as the benchmark Einasto~\cite{Stadel:2008pn,Navarro:2008kc} and NFW~\cite{Navarro:1996gj} models.
The NFW dark matter halo model is the simplest  model consistent with N-body simulations, though there is some scatter about the exact shape of this profile.
The Einasto profile gives a less cuspy profile, as is indicated by higher-resolution simulations.
Due to the uncertain nature of the dark matter profile of the Milky Way, both dark matter profiles are consistent with current observations, so we consider both standard profiles for our Galactic center analyses.
However, the details of the HAWC response are such that the Galactic center analysis is largely independent of the choice of dark matter profile at the Galactic center.
For other dark matter sources, including the dwarf spheroidals and extragalactic sources, our analysis is less dependent on the shape of the inner profile and therefore is largely independent of halo shape as well.
The Einasto profile is given by~\cite{Stadel:2008pn,Navarro:2008kc}
\begin{equation}
\rho_{\rm Einasto}(r)=\rho_s
\exp\left[-\frac{2}{\alpha}\left(\left(\frac{r}{r_s}\right)^{\alpha}
    -1\right)\right]\enspace,
\end{equation}
while the NFW profile is of the form~\cite{Navarro:1996gj}
\begin{equation}
\rho=\frac{\rho_{s}}{(r/r_{s})(1+r/r_{s})^{2}}\enspace,
\end{equation}
where $\rho_{s}$ is the scale density of the profile, $r_s$ is the scale radius of the profile, and in the Einasto profile $\alpha$ parameterizes the profile cuspiness.
For the dwarf galaxies Draco and Coma Berenices, the dark matter profiles are done using an NFW profile with parameters given by~\cite{Abdo:2010ex}.
The Segue 1 dark matter profile is chosen as an Einasto profile with $\alpha=0.3$, from Ref.~\cite{Aliu:2012ga}.
For M31, we use the NFW profile of Ref.~\cite{Geehan:2005aq}.
For the smooth Virgo cluster profile without substructure, we use the smooth NFW profile from Ref.~\cite{Han:2012uw}.
For the substructure-boosted M31 and Virgo cluster profiles, we use the boost factors of $\sim15$ and $\sim35$, respectively, from Ref.~\cite{Sanchez-Conde:2013yxa}.
For the GC, the dark matter profiles considered are the NFW profile and the Einasto profile with $\alpha=0.22$, from Ref.~\cite{Abazajian:2010sq}.
The scale radii, scale density, and distance to the considered sources are shown in Table~\ref{haloparams}.

The dark matter profiles tend to give extended emission across the sky; the GC with an Einasto profile, for instance, extends to $\pm 7.5\degree$ from the GC.
To account for the extended nature of these sources, we performed an optimal angular binning on the sky for each source, maximizing the signal with respect to background within the angular bin, accounting for the HAWC point-spread function.
This binning is discussed in detail in Section~\ref{DataSim}.

\subsection{Calculation of Dark Matter Spectra}
To calculate the photon spectrum for a particular WIMP annihilation channel, we use {\sc PYTHIA 6.4} to simulate the photon radiation of charged particles as well as decays of particles such as the $\pi^0$~\cite{Sjostrand:2006za}, following the method from section 3.2 of Ref.~\cite{Abazajian:2011ak}.
For each final state and each value of $M_\chi$, we calculate the average number of photons in each energy bin per annihilation event, $dN_\gamma/dE$.

Different dark matter models can be dominated by either hadronic, leptonic, or bosonic annihilation channels, so we consider all of these here.
Due to the available phase space, dark matter will usually annihilate into the heaviest available channel.
For this reason we consider the hadronic $t\bar{t}$ and leptonic $\tau^+\tau^-$ channels here.
The $b\bar{b}$ annihilation channel has been studied by several experiments, so we consider that here as well.
A bosonic $W^+W^-$ annihilation channel, motivated by supersymmetric Winos, is the bosonic channel we study.
Finally, dark matter models which are dominated by annihilation to $\mu^+\mu^-$ may be able to explain measured excesses of local positrons~\cite{Adriani:2008zr,FermiLAT:2011ab,Aguilar:2013qda}, so we also analyze that channel.

In addition to the prompt gamma-ray emission discussed above, each annihilation channel also produces many charged particles (protons, anti-protons, electrons, and positrons).
As these charged particles propagate, they can undergo inverse Compton (IC) scattering off low-energy background photons from starlight, the infrared background, and the cosmic microwave background (CMB).
These IC-scattered photons can be measured in addition to the prompt gamma-ray emission.
Particularly for leptonic dark matter annihilation modes, which produce few prompt photons, the addition of this IC emission can increase the dark matter gamma-ray flux.
Because the IC emission usually peaks at lower energy than the prompt emission, this emission particularly affects the limits on the highest dark matter masses.
Also, because the IC emission from multi-TeV-mass dark matter annihilation extends the gamma-ray spectrum down to much lower energies, the inclusion of IC emission greatly improves the multi-TeV dark matter limits from lower-energy experiments; we encourage these experiments to consider IC emission in their dark matter analyses to increase their sensitivity to higher dark matter masses.

For our calculation of the IC emission from dark matter annihilations, we consider only the IC from electrons and positrons on the CMB.
While considering IC from higher-energy photon fields could improve these limits further, we only consider the model-independent CMB limit here to be conservative.
Similarly, we also do not include the bremsstrahlung emission from emitted electrons or positrons discussed in Ref.~\cite{Cirelli:2013mqa}.
We calculate the IC component of our spectra using the diffusion-free approximation from Refs.~\cite{Profumo:2009uf,Cirelli:2009vg,Abazajian:2010zb}.
In this framework, we obtain the annihilation electron+positron spectrum similarly to the prompt gamma-ray emission above, using {\sc PYTHIA 6.4}.
These are then scattered with CMB photons to produce the IC gamma-ray spectrum, which is added to the prompt gamma-ray spectrum.

While the diffusion-free approximation is not exact, the average path-length for a TeV electron is less than 1 kpc~\cite{Papucci:2009gd} for galactic systems and a diffusion-free model is a reasonable approximation of the IC emission for multi-GeV and TeV photons in such systems~\cite{Pieri:2009je}.
In regions where the electron injection rate varies on scales much less than the $\sim1$ kpc path-length, such as the Galactic center, this approximation breaks down and a full simulation including diffusion would be needed~\cite{Pieri:2009je}, so we do not show the GC limits here including IC component. Also, for objects with scale radii on the order of the electron diffusion path-length, the IC emission would have spatial extent based on the details of the particle diffusion. Therefore, for the dwarf galaxies, we do not show IC limits from dwarf galaxies here.

\section{Data Simulation and Analysis}~\label{DataSim}
\subsection{Simulation of HAWC}
Approximately 1.2 billion gamma-ray, 1.1 billion proton, 900 million helium, and 60 million heavier nuclei induced air showers were simulated using CORSIKA v6990~\cite{citeulike:9245764} with FLUKA version 2011~\cite{FLUKA1,FLUKA2} and QGSJET II.
Each primary was drawn from an $E^{-2}$ spectrum to optimize for high statistics at high energies without spending too much time processing high-energy events.
The resulting showers were then injected into a detailed detector response simulated within GEANT4 9.5.p01~\cite{citeulike:6822207,citeulike:9339116}.
The GEANT4 portion of the simulation, originally developed for Milagro~\cite{citeulike:9259094}, includes detailed descriptions of the geometrical and optical properties of the HAWC tanks and photomultiplier tubes (PMTs).

The simulated HAWC detector consists of 300 steel tanks, each tank is 7.3\,m in diameter and 5.4\,m tall with 4.5\,m of water above each of 4 PMTs.
Each tank contains three 8-inch Hamamatsu R5912 PMTs arrayed around a central 10-inch Hamamatsu R7081 PMT with a high quantum efficiency photocathode.
The Hamamatsu R5912 PMTs are inherited from the Milagro experiment, so the simulations benefit from previously constructed GEANT4 models \cite{citeulike:9259208}.

The output from the GEANT4 portion of the simulation was then reconstructed using the HAWC reconstruction software, yielding the shower properties which were then used to characterize the detector sensitivity to particular sources.
The characterization of the sensitivity of HAWC to a particular source was similar to the analysis of measured HAWC data:
\begin{itemize}
  \item Simulated hadrons are weighted to reproduce observed cosmic-ray spectra of Ref.~\cite{Yoon:2011aa}.
  \item Simulated gammas are weighted to reproduce expected gamma-ray spectrum from a source.
  \item Energy binning is done based on the number of hit PMTs across the array (nHit), independent of the source spectrum or declination.
  \item Source and background rates are calculated per nHit bin for each source spectrum and declination. The declination dependence of the HAWC sensitivity is shown figure 2 of Ref.~\cite{Abeysekara:2013tza}.
  \item The significance is calculated assuming gaussian statistics are valid for the given observation period.
\end{itemize}
Details of this process can be found in Ref.~\cite{Abeysekara:2013tza}.
\subsection{Analysis Verification}
This analysis differs from the HAWC point-source analysis~\cite{Abeysekara:2013tza} only in that it allows for much angular larger bins for known source morphologies.
The optimal angular bins were found by convolving each theoretical DM gamma-ray source morphology with the point spread function found from the above simulations and chosing the area around the source which maximizes the signal-to-noise ratio for each nHit bin (see Ref.~\cite{Abeysekara:2013tza} for further details on HAWC binning).

The GC is a fairly extended source near the edge of HAWC's field-of-view where the trigger rate due hadron-induced air showers is changing rapidly. Air showers near the edge of the HAWC field-of-view must pass through more atmosphere than those coming from near zenith. Therefore, the simulated background rate for these showers is more dependent on the CORSIKA atmospheric model than those from zenith. The cosmic ray rate over the declinations covered by the GC optimal spatial bin changes by as much as 30 percent. Therefore, the changing background rate over the GC morphology must contribute strictly less than a 30 percent systematic uncertainty in sensitivity. The HAWC sensitivity dependence with declination is shown figure 2 of Ref.~\cite{Abeysekara:2013tza}

There is also a possible systematic uncertainty due to possible gamma-ray source contamination. For our considered sources, only the Galactic center and the Virgo cluster have known gamma-ray sources which could degrade our sensitivity: the Galactic plane and GC itself for the GC analysis and M87 for the Virgo cluster analysis. To robustly account for the systematic errors from such sources, we masked out the inner $0.5\degree$ around the gamma-ray sources from our analysis to determine the change in our dark matter limits. The $0.5\degree$ region is larger than the HAWC point-spread function at the energies relevant for this analysis. With the possible point-sources removed, our GC limits weakened by 13\% for the NFW profile. The dark matter limits from the smooth Virgo cluster are weakened by 32\%. The limits from the GC with an Einasto profile and the Virgo cluster extended by dark matter substructure are not significantly affect by the removal of the inner part of the profile. We demonstrate our combined systematics from possible point-source contamination and the background rate near the edge of the HAWC field-of-view in Figures~\ref{HAWCM31Virgo} and~\ref{HAWCGC} as gray bands around each line, to the right of the plots.

\section{Projected Dark Matter Limits from HAWC}
\begin{figure*}
\begin{center}$
\begin{array}{c}
\begin{array}{cc}
\includegraphics[width=0.5\textwidth]{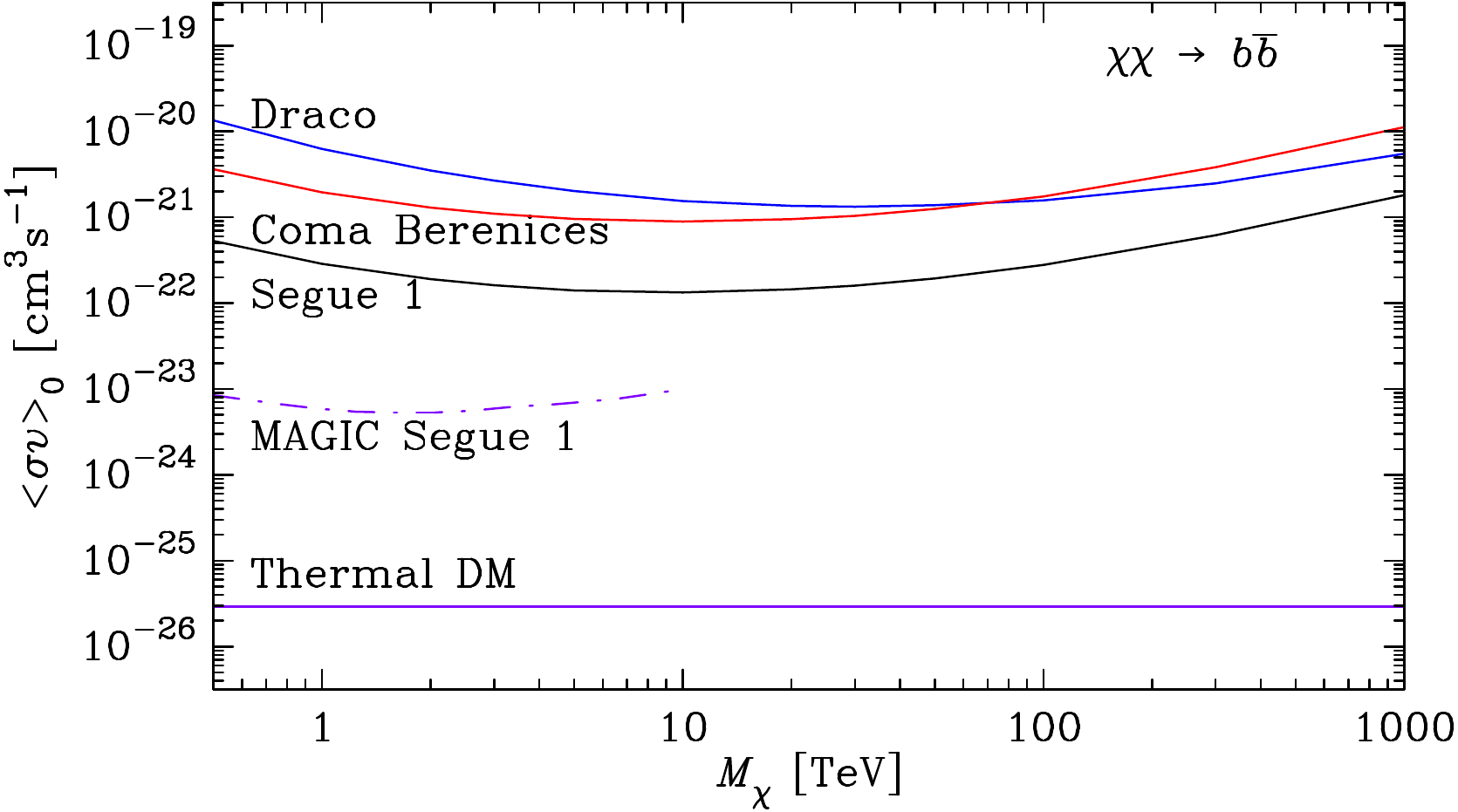} &
\includegraphics[width=0.5\textwidth]{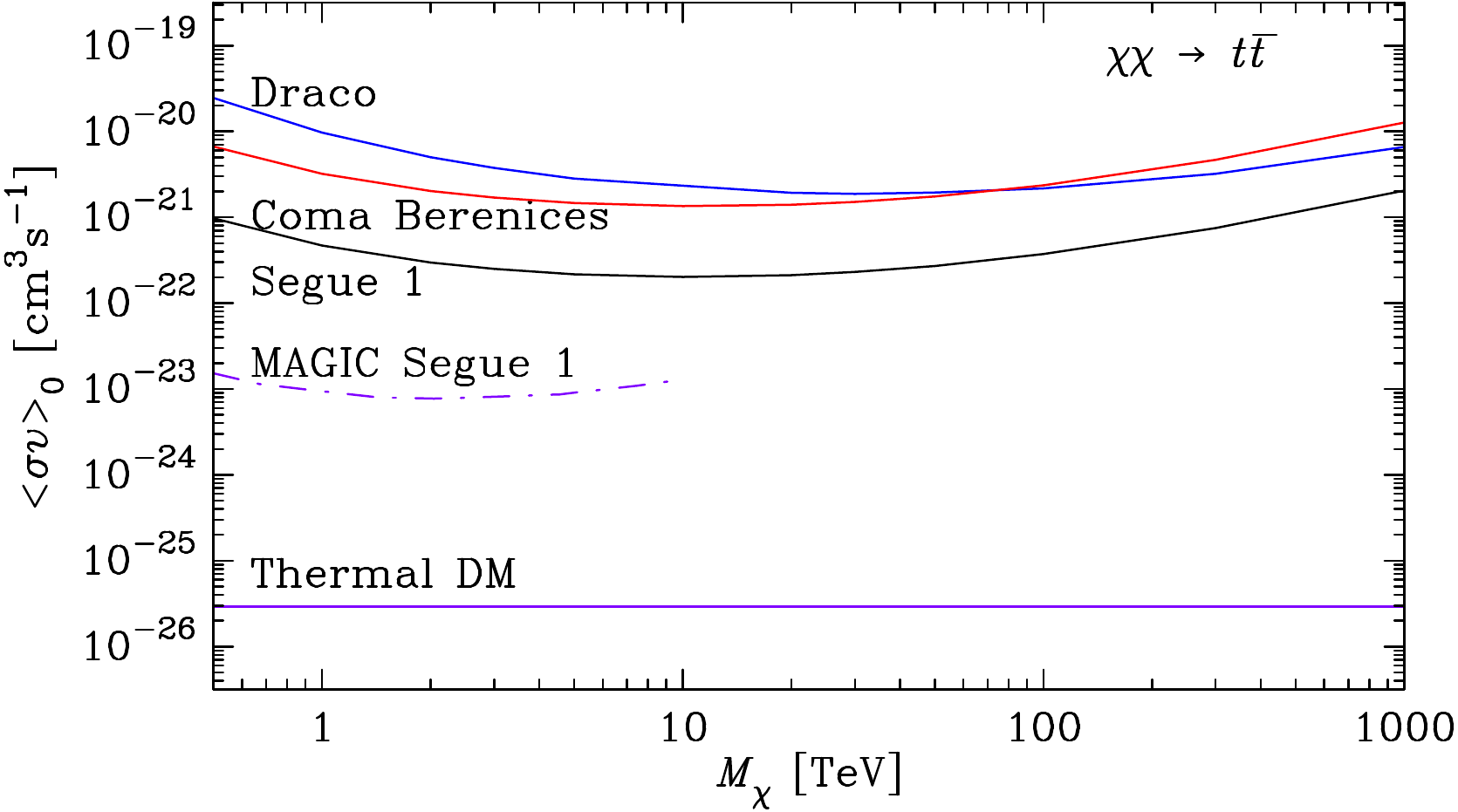}
\end{array}\\
\begin{array}{cc}
\includegraphics[width=0.5\textwidth]{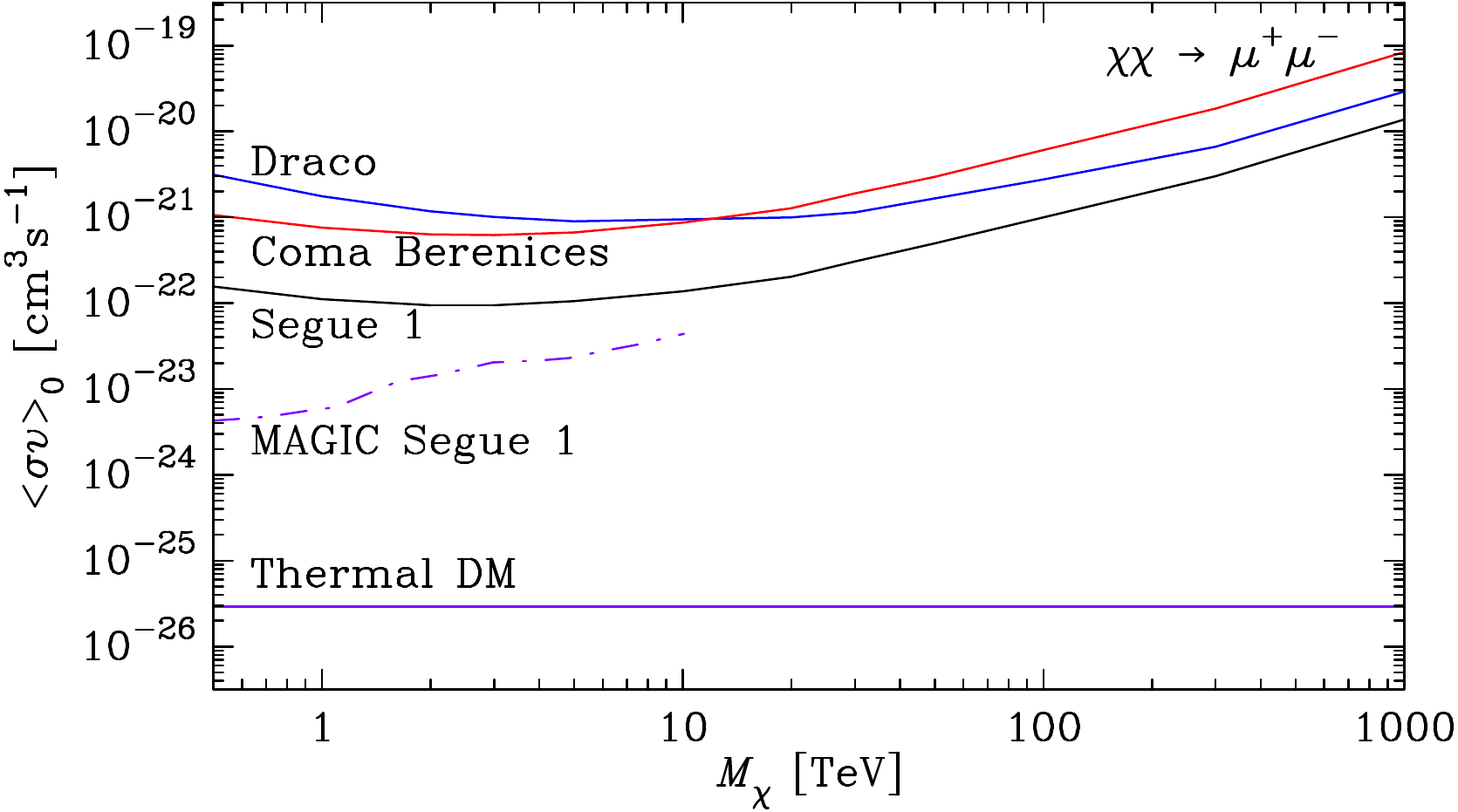} &
\includegraphics[width=0.5\textwidth]{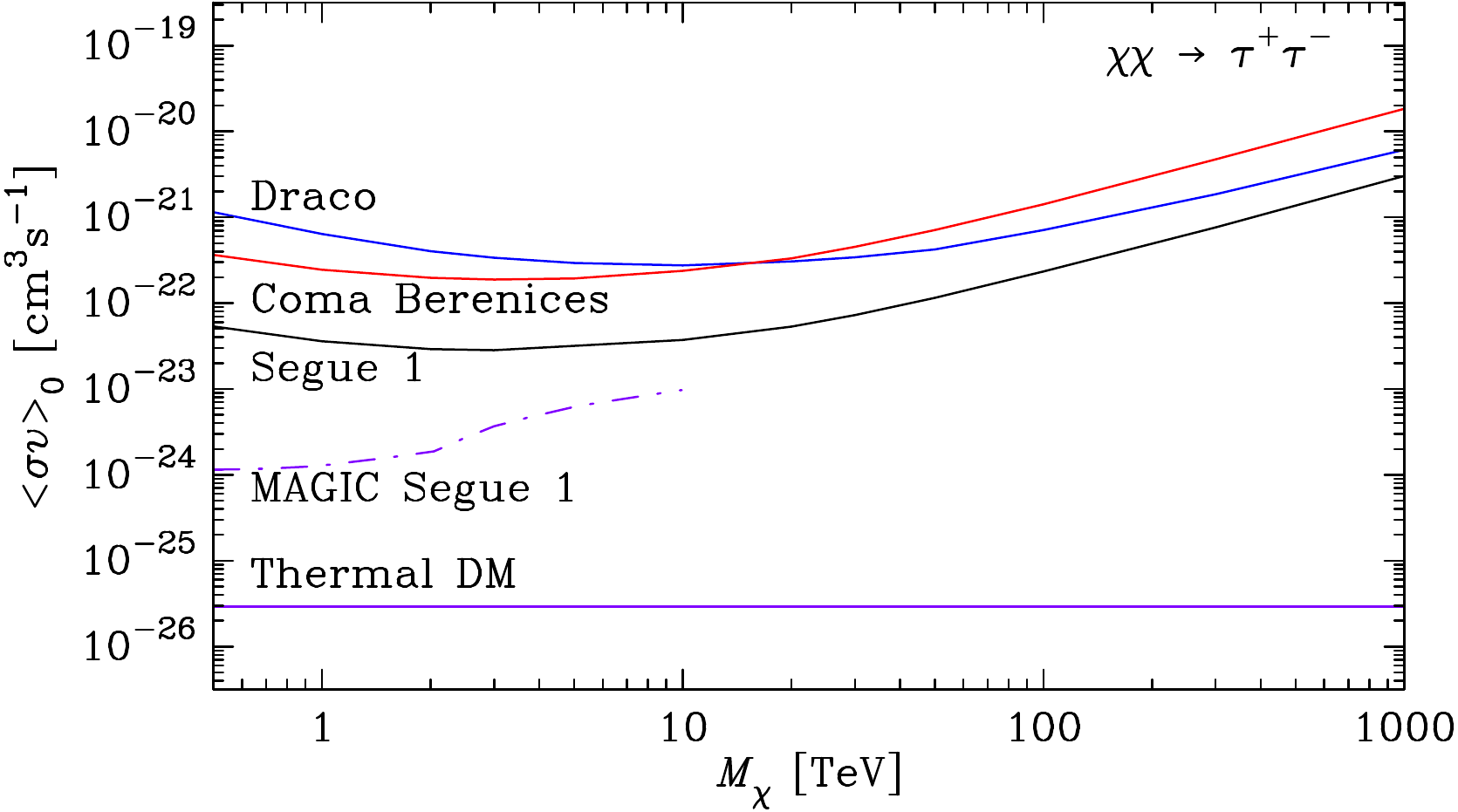}
\end{array}\\
\begin{array}{c}
\includegraphics[width=0.5\textwidth]{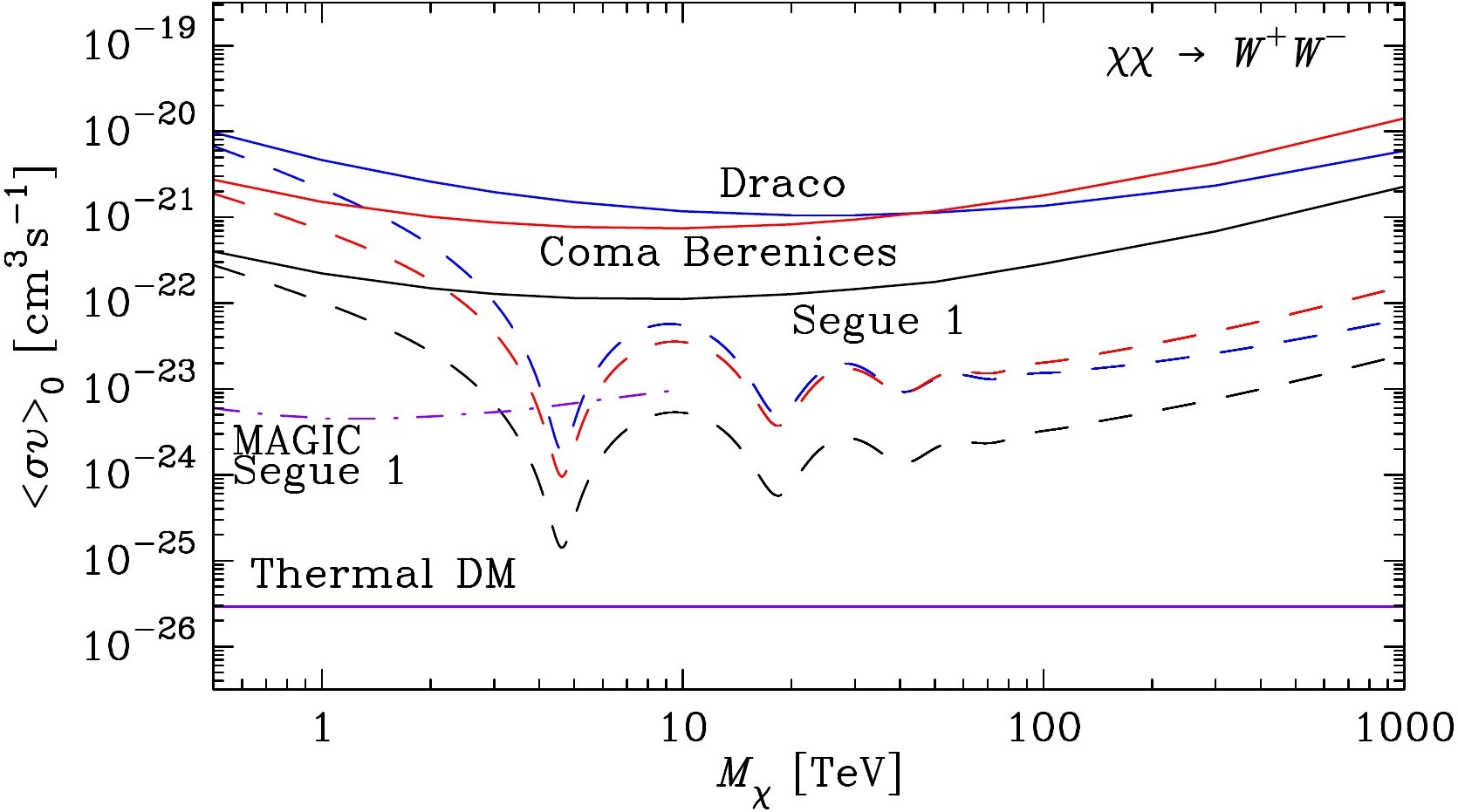}
\end{array}
\end{array}$
\end{center}
\caption[The projected dark matter limits from dwarf galaxies for HAWC after five years.]{\small The projected dark matter limits from dwarf galaxies for HAWC after five years, for the $b\bar{b}$, $t\bar{t}$, $\mu^+\mu^-$, $\tau^+\tau^-$, and $W^+W^-$ dark matter annihilation channels.
From top to bottom, the curves are for Draco (blue), Coma Berenices (red), and Segue 1 (black).
The solid curves are the dark matter limits for just the prompt gamma-ray emission.
In the $W^+W^-$ plot, the dashed curves are the limit on the early-universe annihilation cross-section when natural Sommerfeld enhancement is included in the cross-section today (with $v_{\rm rel}=300\rm\,km\,s^{-1}$).
We show the 157.9-hour MAGIC dark matter exclusion limits from Segue 1 as the purple dot-dashed curve for comparison~\cite{Aleksic:2013xea}.
Note that Sommerfeld enhancement improves the MAGIC results similarly to those of HAWC.
The solid purple line shows the expected dark matter thermal cross-section.
All limits are at 95\% CL.
\label{HAWCdwarfs}}
\end{figure*}

\begin{figure*}
\begin{center}$
\begin{array}{c}
\begin{array}{cc}
\includegraphics[width=0.5\textwidth]{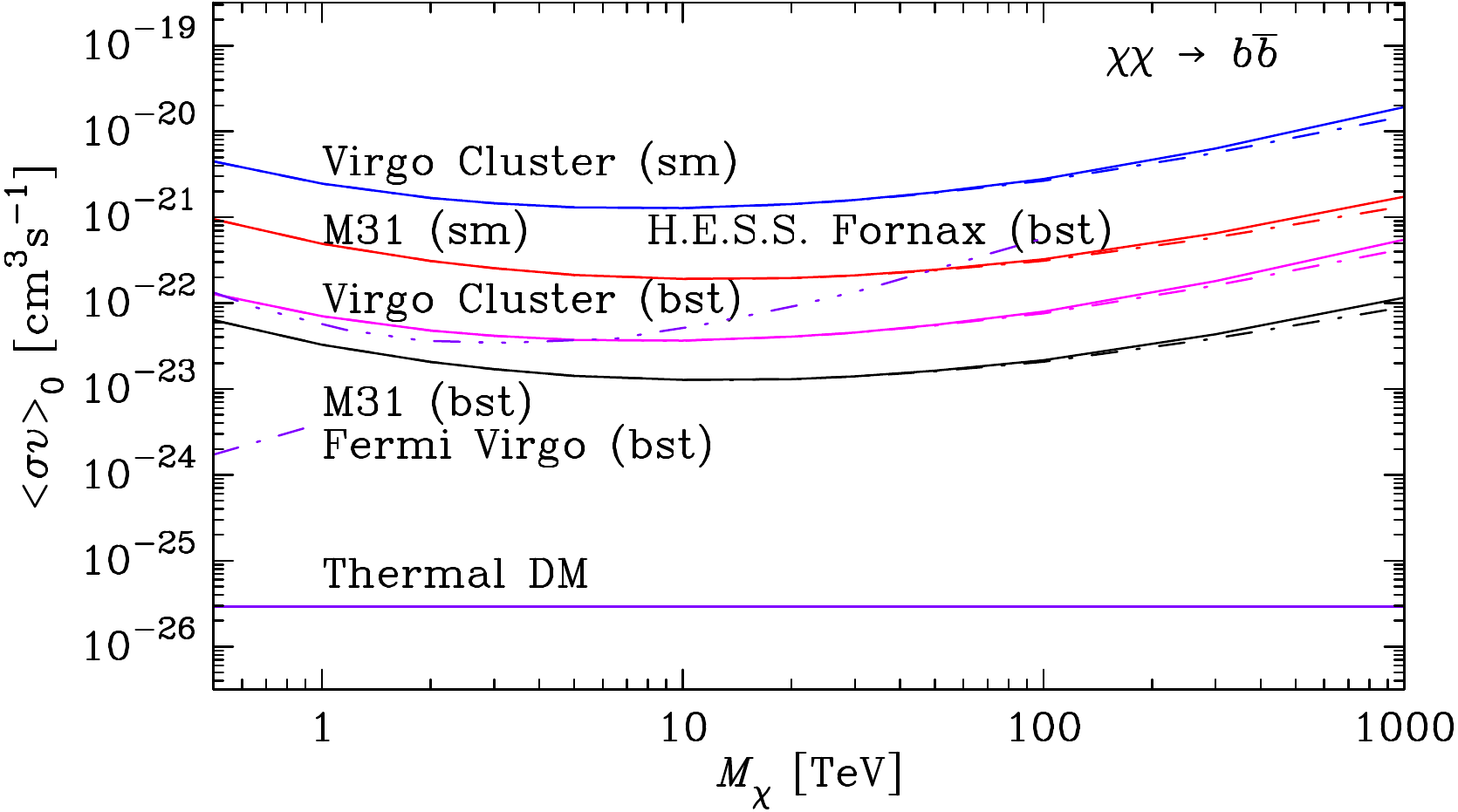} &
\includegraphics[width=0.5\textwidth]{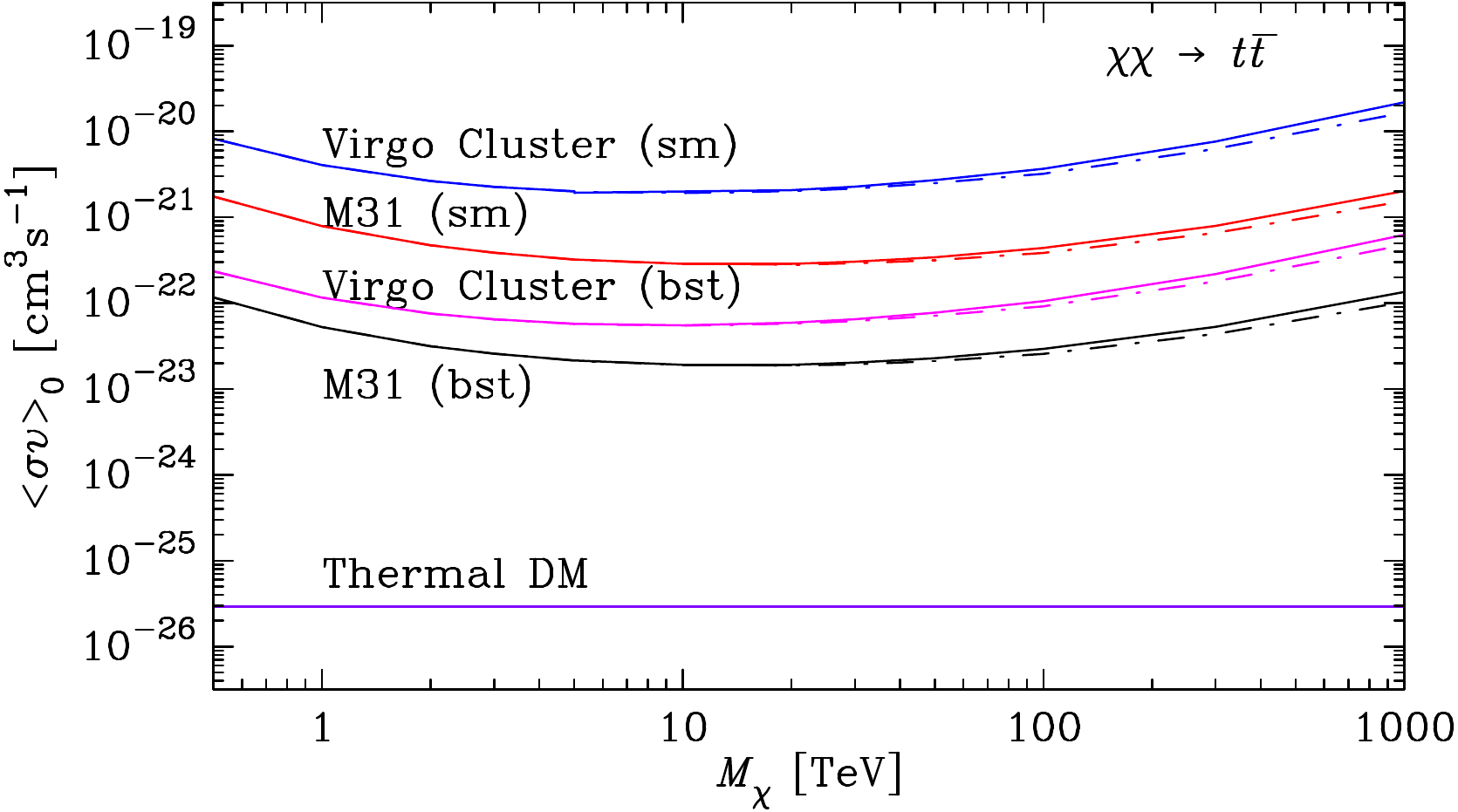}
\end{array}\\
\begin{array}{cc}
\includegraphics[width=0.5\textwidth]{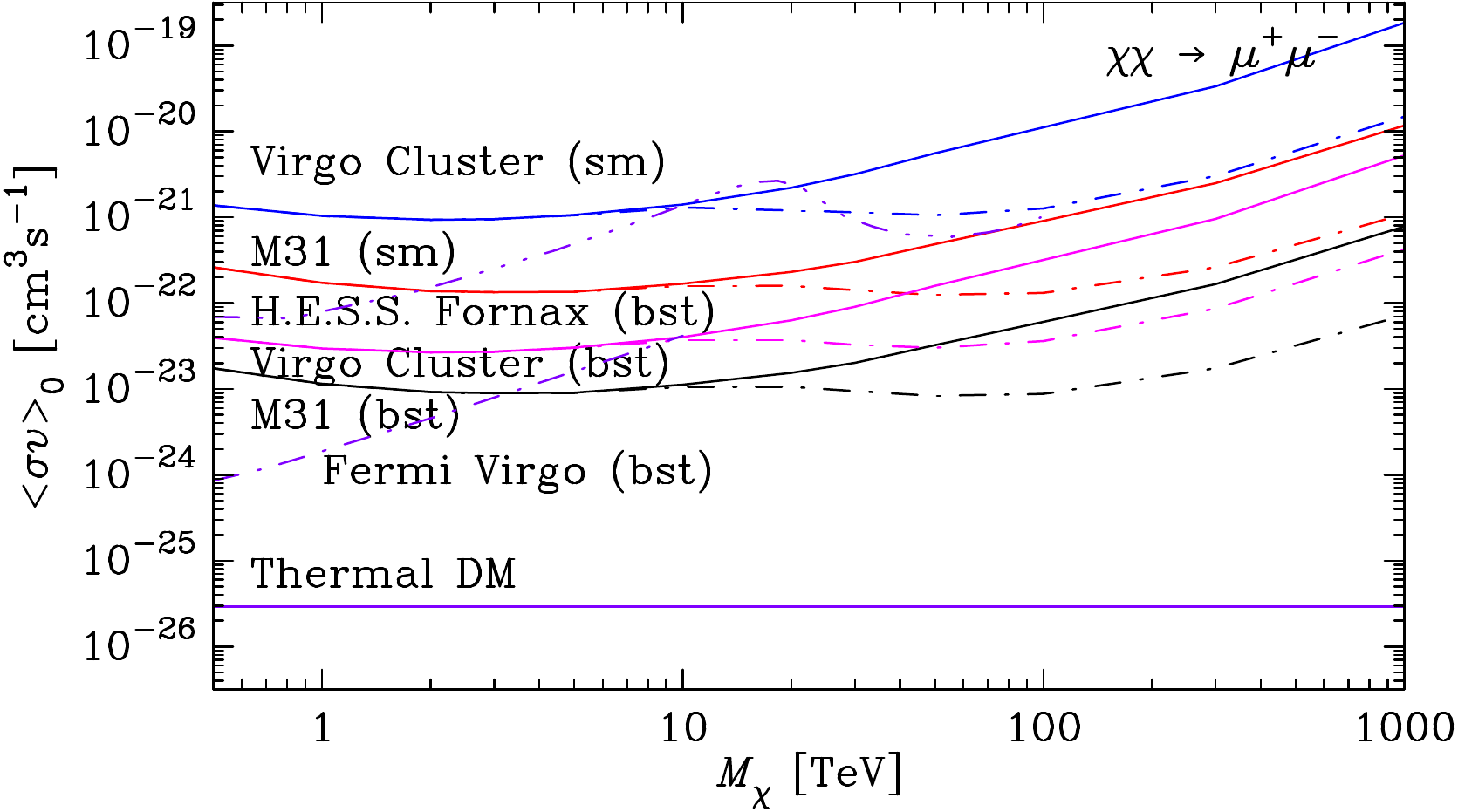} &
\includegraphics[width=0.5\textwidth]{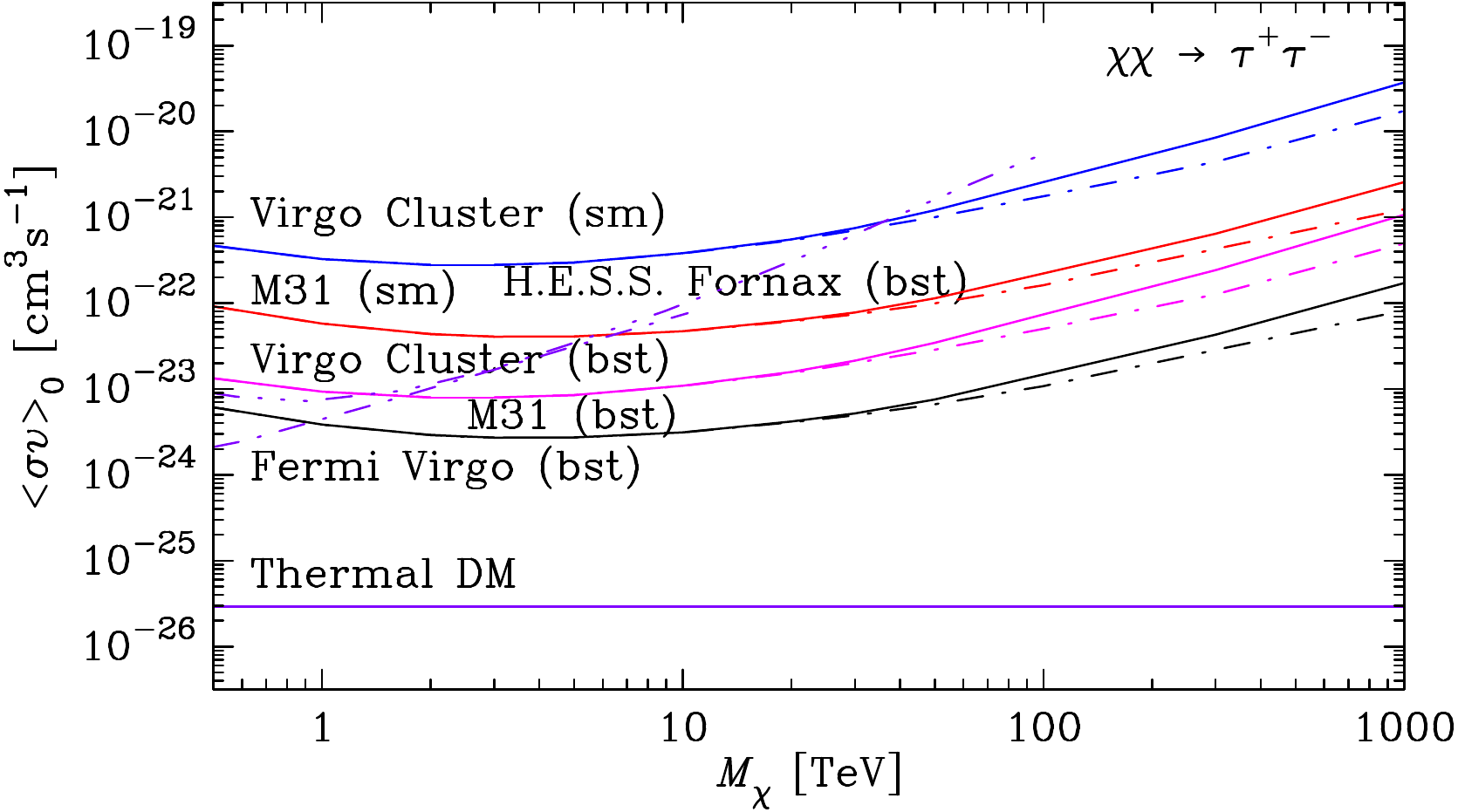}
\end{array}\\
\begin{array}{c}
\includegraphics[width=0.5\textwidth]{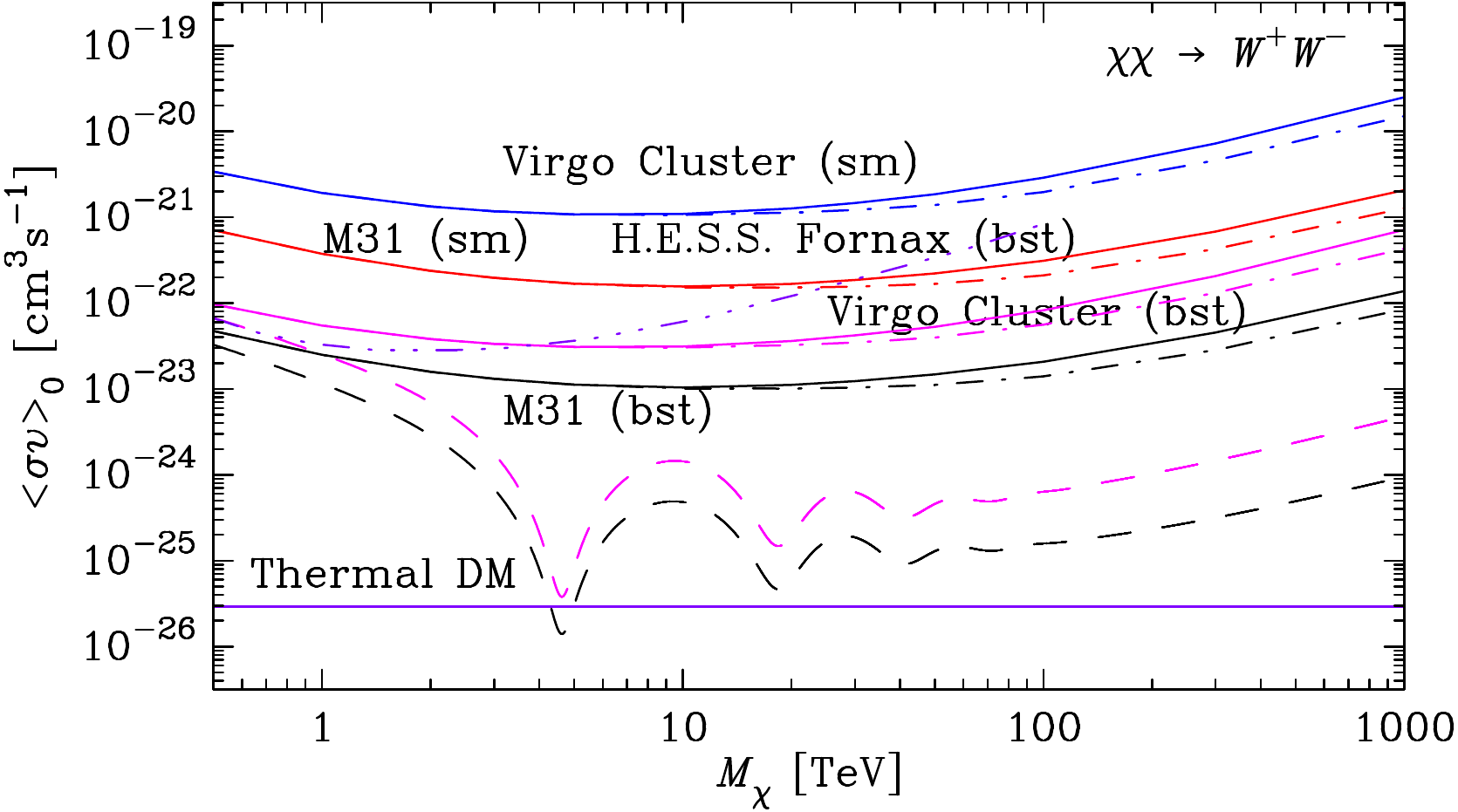}
\end{array}
\end{array}$
\end{center}
\caption[The projected dark matter limits from the Virgo cluster and the galaxy M31 for HAWC after five years.]{\small The projected dark matter limits from the Virgo cluster and the galaxy M31 for HAWC after five years, for the $b\bar{b}$, $t\bar{t}$, $\mu^+\mu^-$, $\tau^+\tau^-$, and $W^+W^-$ dark matter annihilation channels.
From top to bottom, the curves are for the Virgo cluster with a smooth (sm) NFW profile (blue), M31 with a smooth (sm) NFW profile (red), the substructure-boosted (bst) Virgo cluster (magenta), and the substructure-boosted (bst) M31 (black).
The triple-dot-dashed purple line is the limit from the H.E.S.S. observatory observations of the Fornax cluster~\cite{Abramowski:2012au}, boosted (bst) using the substructure boost model of Ref.~\cite{Sanchez-Conde:2013yxa}.
The dot-dashed purple line is the limit from the Fermi-LAT observations of the Virgo cluster~\cite{Han:2012uw}, boosted (bst) using the substructure boost model of Ref.~\cite{Sanchez-Conde:2013yxa}.
For the $\mu^+\mu^-$ channel, both the H.E.S.S. Fornax limit and the Fermi-LAT Virgo limit are for a combination of prompt emission and IC emission.
Here we employ the substructure boost of 35 for the Virgo cluster, 15 for M31, and 29 for the Fornax cluster, based on Ref.~\cite{Sanchez-Conde:2013yxa}.
The solid curves are the dark matter limits for just the prompt gamma-ray emission, and the dot-dashed curves are the limits considering both the prompt gamma-ray mission and the IC emission from electrons and positrons scattering on the CMB.
In the $W^+W^-$ plot, the dashed curves are the limit on the early-universe annihilation cross-section when natural Sommerfeld enhancement is included in the cross-section today (with $v_{\rm rel}=300\rm\,km\,s^{-1}$).
The width of the gray bands above the smooth Virgo cluster lines to the right of the figure indicate the range in the dark matter limit for all masses due to possible uncertainty from to point-source subtraction in the analysis.
The solid purple line shows the expected dark matter thermal cross-section.
All limits are at 95\% CL.
\label{HAWCM31Virgo}}
\end{figure*}

\begin{figure*}
\begin{center}$
\begin{array}{c}
\begin{array}{cc}
\includegraphics[width=0.5\textwidth]{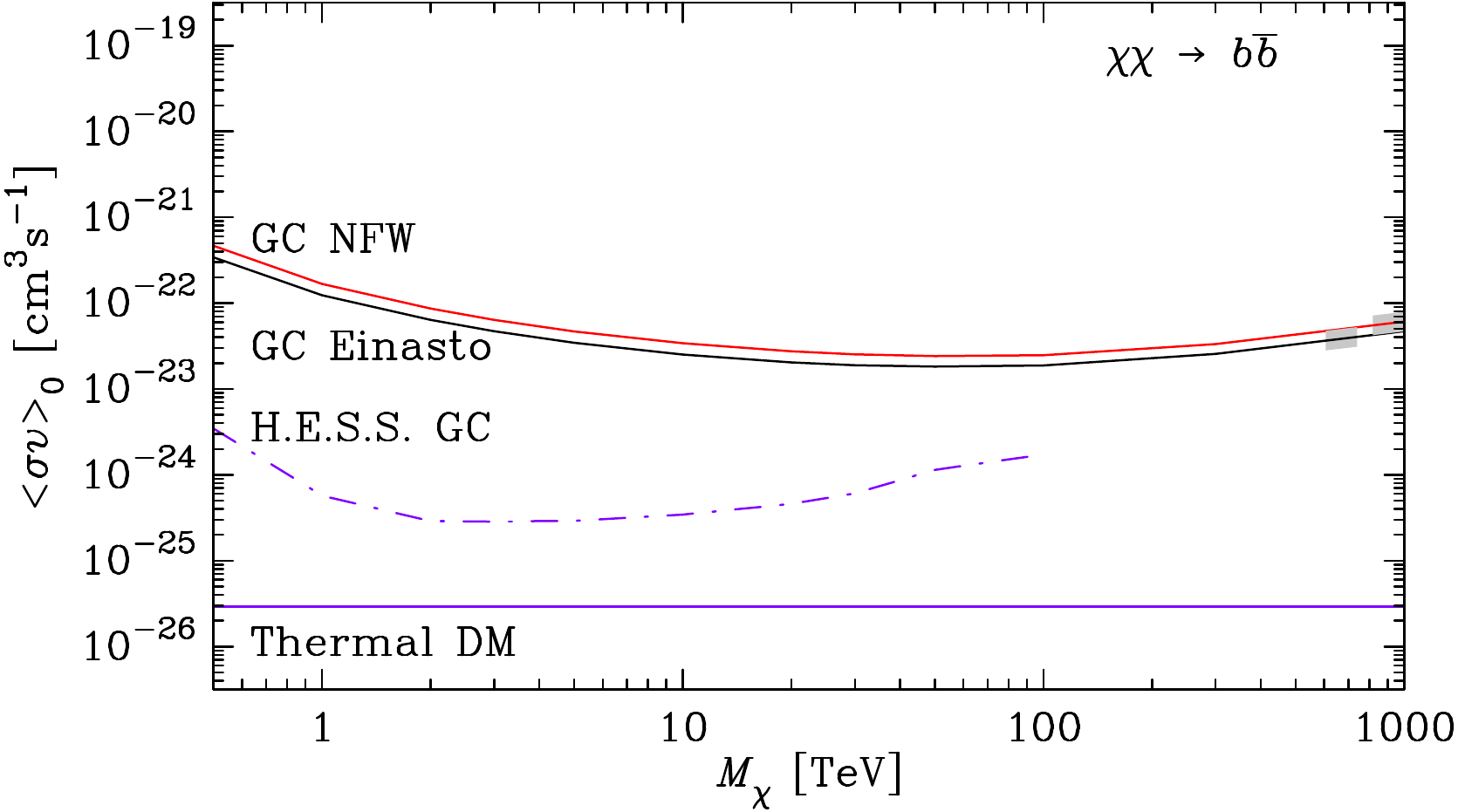} &
\includegraphics[width=0.5\textwidth]{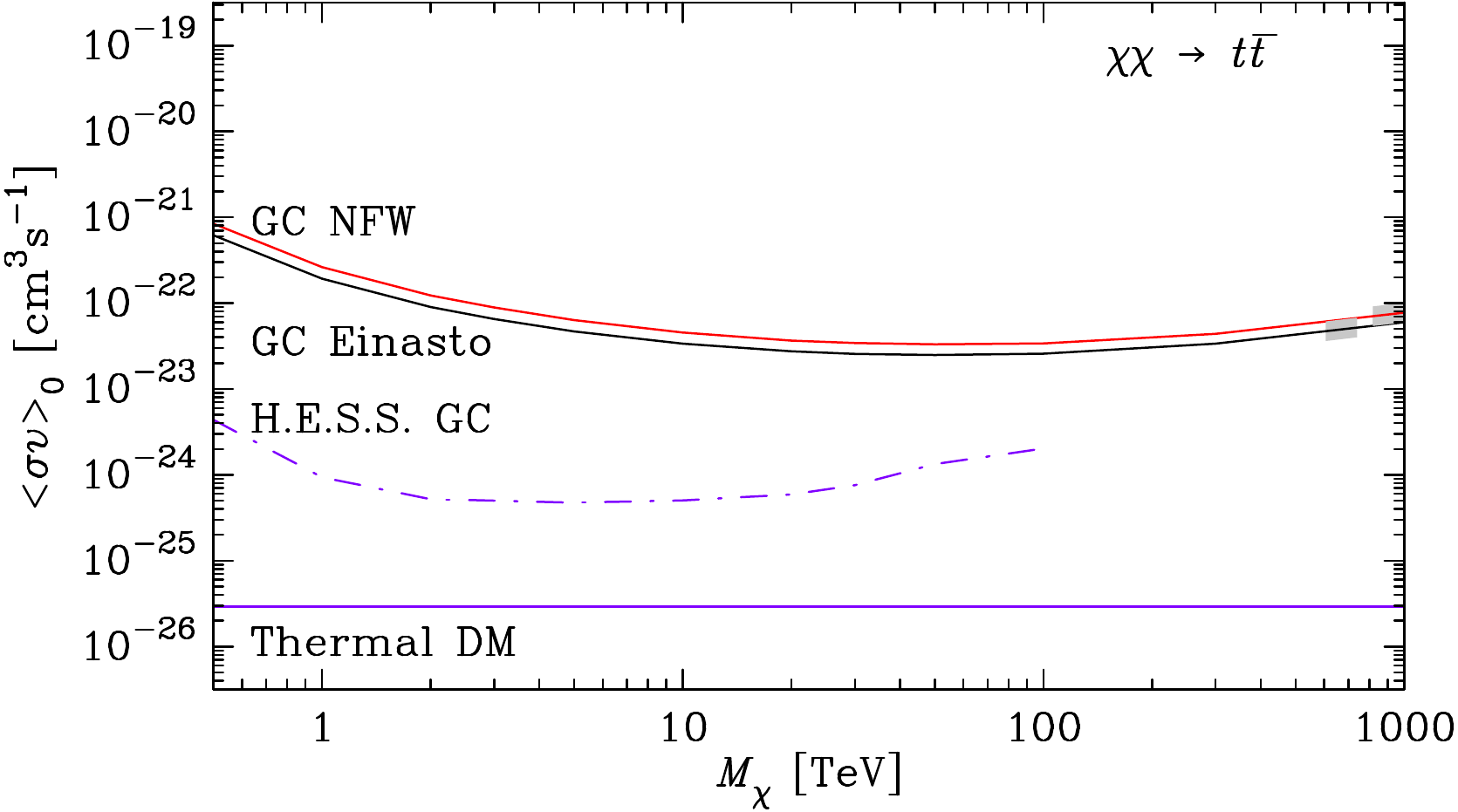}
\end{array}\\
\begin{array}{cc}
\includegraphics[width=0.5\textwidth]{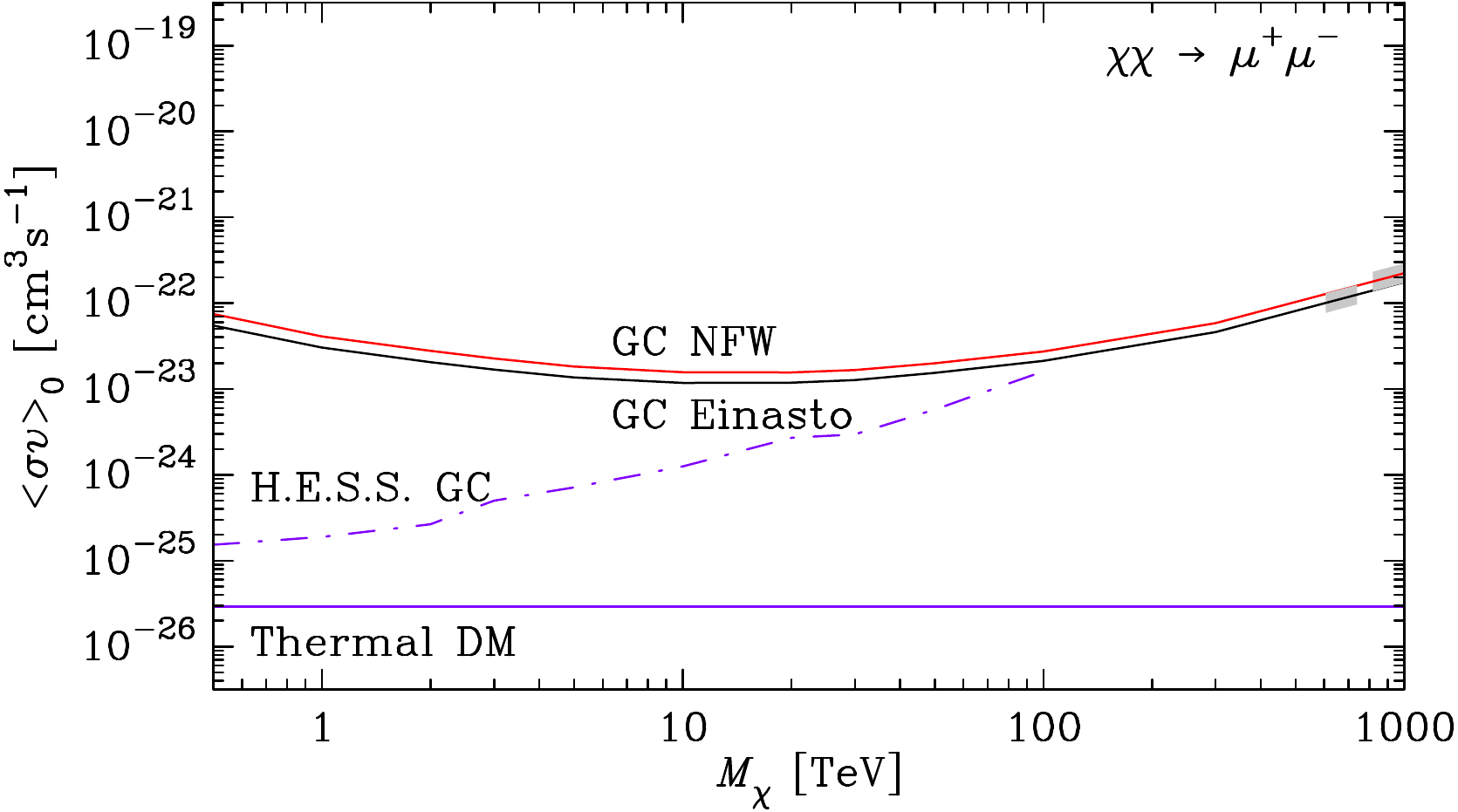} &
\includegraphics[width=0.5\textwidth]{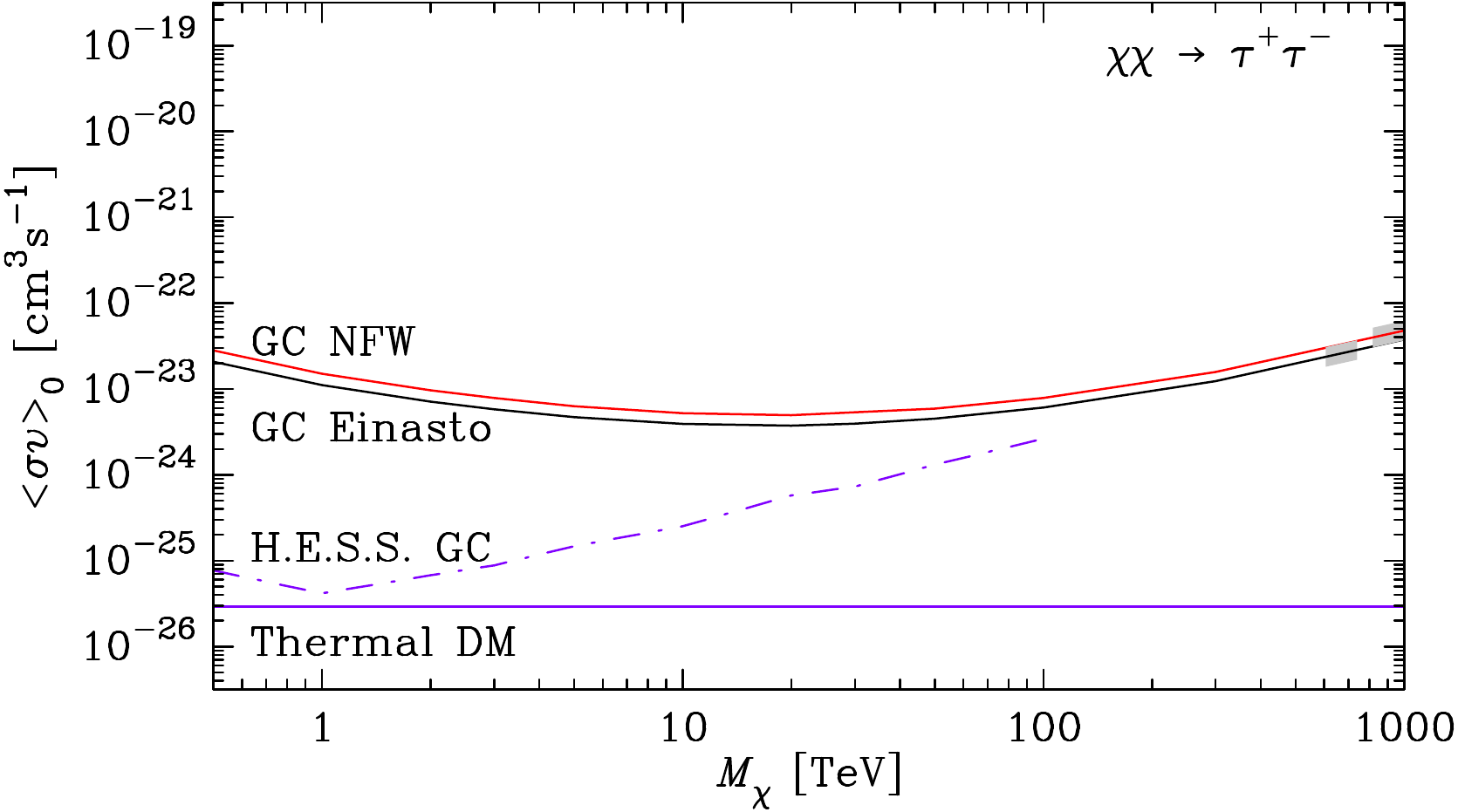}
\end{array}\\
\begin{array}{c}
\includegraphics[width=0.5\textwidth]{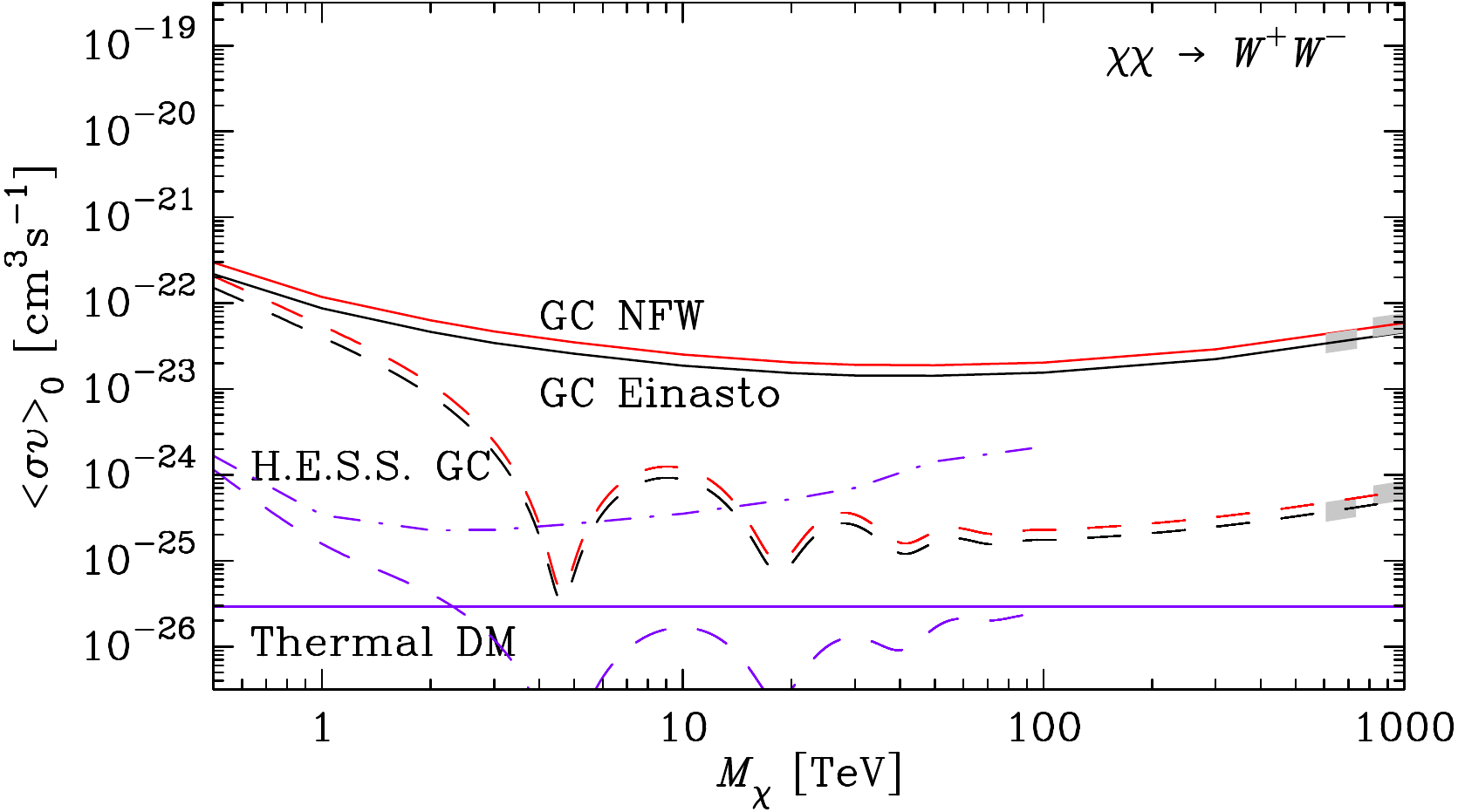}
\end{array}
\end{array}$
\end{center}
\caption[The projected dark matter limits from the GC for HAWC after five years.]{\small The projected dark matter limits from the GC for HAWC after five years, for the $b\bar{b}$, $t\bar{t}$, $\mu^+\mu^-$, $\tau^+\tau^-$, and $W^+W^-$ dark matter annihilation channels.
  From top to bottom, the curves are for the GC assuming an NFW profile (red) and assuming an Einasto profile (black).
  The solid curves are the dark matter limits for just the prompt gamma-ray emission, and the dot-dashed curves are the limits considering both the prompt gamma-ray mission and the IC emission from electrons and positrons scattering on the CMB.
In the $W^+W^-$ plot, the dashed curves are the limit on the early-universe annihilation cross-section when natural Sommerfeld enhancement is included in the cross-section today (with $v_{\rm rel}=300\rm\,km\,s^{-1}$).
The width of the gray bands on the right of the figure indicate the range in the dark matter limit for all masses due to the combined uncertainty from the HAWC sensitivity near the edge of its field-of-view and the possible uncertainty due to point-source subtraction in the analysis.
We show the H.E.S.S. 112-hour dark matter exclusion limits from the GC as the purple dot-dashed curve for comparison and in the $W^+W^-$ plot, the H.E.S.S. limit with Sommerfeld enhancement is shown as the purple dashed curve~\cite{Abramowski:2011hc,Abazajian:2011ak}.
The solid purple line shows the expected dark matter thermal cross-section.
All limits are at 95\% CL.
\label{HAWCGC}}
\end{figure*}

Through detailed simulation of the HAWC gamma-ray sensitivity and backgrounds~\cite{Abeysekara:2013tza}, we have determined the significance of the dark matter flux for five annihilation channels, a range of dark matter masses, and several different dark matter sources.
Assuming that no dark matter signal is observed above background, we convert the source significance into exclusion curves of the dark matter cross section for given dark matter mass.
In Figure~\ref{HAWCdwarfs}, the curves are the projected 95\% CL limits from the Draco dwarf galaxy (blue curves), the Coma Berenices dwarf galaxy (red curves), and the Segue 1 dwarf galaxy (black curves).
In Figure~\ref{HAWCM31Virgo}, the curves are the projected 95\% CL limits from the M31 galaxy with a smooth NFW profile (red curves), the Virgo cluster with a smooth NFW profile (blue curves), the substructure-boosted Virgo cluster (magenta curves), and the substructure-boosted M31 (black curves).
In Figure~\ref{HAWCGC}, the curves are the projected 95\% CL limits from the GC assuming either an NFW profile (red curves) or an Einasto profile (black curves).
For our exclusion curves, we have assumed a five-year observation time for HAWC and derived the 95\% confidence-level (CL) cross-section limits WIMPs which annihilate with a 100\% branching ratio into $b\bar{b}$, $t\bar{t}$, $\mu^+\mu^-$, $\tau^+\tau^-$, or $W^+W^-$ annihilation channels.

In each plot, the solid curves show the limits from only the prompt gamma-ray emission from the sources, while the dot-dashed lines show the limits when both the prompt and IC emission are included.
For the he hadronic $b\bar{b}$ and $t\bar{t}$ channels and the bosonic $W^+W^-$ channel, the addition of the IC emission to the prompt gamma-ray emission only weakly improves the dark matter limits.
In contrast, the leptonic $\tau^+\tau^-$ channel benefits from the inclusion of the IC spectra around 100\,TeV, with an improvement of its limits by a factor of $\sim2$ for high mass.
The leptonic $\mu^+\mu^-$ channel benefits the most from the IC emission, with significantly-improved limits above 10\,TeV and an order of magnitude improvement at high masses.
Our figures show the limits both with and without the IC component for comparison to other experiments' limits.
Additionally, the IC component shown only includes the upscattering of CMB photons, not starlight or the infrared background, and therefore should be more constraining when these additional components are considered.

In the $W^+W^-$ annihilation channel plots of Figures~\ref{HAWCdwarfs}-\ref{HAWCGC}, the dashed lines indicate the limit when natural Sommerfeld enhancement from the exchange of Standard Model $W$ gauge bosons is included, for $v_{\rm rel}=300\rm\,km\,s^{-1}$.
F ease of comparison with other experiments, we show the limit without the Sommerfeld enhancement (solid line) as well.
The inclusion of natural Sommerfeld enhancement for the bosonic $W^+W^-$ annihilation channel improves the limits dramatically (Figures~\ref{HAWCdwarfs}-\ref{HAWCGC}).
For dark matter masses as little as 1\,TeV, the Sommerfeld enhancement improves the limits by a factor of $\sim2$, while at high masses, the limits are improved by 2 orders of magnitude.
Note that the amount of Sommerfeld enhancement increases as the dark matter velocity decreases, so our choice of $v_{\rm rel}=300\rm\,km\,s^{-1}$ here gives only a moderate amount of Sommerfeld enhancement.
Especially in dark matter substructure, including dwarf galaxies, dark matter velocities are expected to be much lower and these limits should improve substantially.
In Figure~\ref{Sommfig}, it can be seen that for $v_{\rm rel}\sim0$, these limits would improve by over an order of magnitude at the highest masses in regions with much lower dark matter velocities.

In Figures~\ref{HAWCdwarfs}-\ref{HAWCGC} one can see that the limits from the hadronic $b\bar{b}$ and $t\bar{t}$ channels and the bosonic $W^+W^-$ channel without Sommerfeld enhancement are similar.
The $\mu^+\mu^-$ channel without the IC component gives a factor of 3 stronger limits at low mass and a factor of 3 worse limits at high mass than the hadronic and bosonic channels.
The  $\mu^+\mu^-$ channel with the inclusion of the IC emission also gives a factor of 3 stronger limits than the hadronic and bosonic channels at low mass, but gives similar limits to the hadronic and bosonic channels at high mass.
The $\tau^+\tau^-$ channel is the un-enhanced channel most strongly constrained by HAWC, giving an order of magnitude stronger limits at low mass and similar constraints at high mass to the hadronic and bosonic channels.
However, above $\sim2$ TeV, the limits on the $W^+W^-$ annihilation channel above are the strongest-constrained HAWC annihilation channel.
\section{Discussion}
\subsection{HAWC Single-source Projected Limits}
Figure~\ref{HAWCdwarfs} demonstrates the HAWC sensitivity to dark matter annihilation in single dwarf spheroidal galaxies.
While the Draco dwarf galaxy has a dark matter $J$-factor that is larger than Coma Berenices by a factor of $\sim 4$, Coma Berenices has a declination culminating closer to the zenith at Sierra Negra than Draco (see Table~\ref{haloparams}).
Therefore, the sensitivity to dark matter annihilations in Coma Berenices is similar to that from Draco, indicating the importance of observational losses due to zenith angle.
As expected, the most favorable targets for HAWC dark matter analyses of dwarf spheroidal galaxies are those which culminate close to zenith for HAWC.
With a zenith angle more favorable than Coma Berenices and a larger dark matter $J$-factor than Draco, however, Segue 1 is the strongest dwarf galaxy candidate for HAWC to observe a dark matter signal.

Dark matter limits from lower-energy studies of dwarf spheroidal galaxies have been used to constrain the dark matter cross-section for masses below tens of TeV.
The Fermi-LAT~\cite{Abdo:2010ex,GeringerSameth:2011iw,Ackermann:2011wa}, H.E.S.S.~\cite{Abramowski:2010aa}, VERITAS~\cite{Acciari:2010ab,Aliu:2012ga}, and MAGIC~\cite{Aleksic:2011jx,Aleksic:2013xea} collaborations have all looked at dwarf spheroidal galaxies for low-background signals of dark matter annihilation.
In Figure~\ref{HAWCdwarfs}.
we show the limit with the strongest constraint for the masses we consider here - the 157.9-hour MAGIC observation of Segue 1~\cite{Aleksic:2013xea}.
Though the MAGIC limits cut off above 10 TeV, the HAWC Segue 1 limits for the hadronic and bosonic channels should be stronger than the MAGIC limits above $\sim 100 {\rm\,TeV}$.
The HAWC limit improves on the MAGIC limit more strongly for the leptonic channels, with the HAWC Segue 1 limit more constraining than the MAGIC limit above $\sim 20 {\rm\,TeV}$.

Figure~\ref{HAWCM31Virgo} demonstrates that extragalactic sources, though far away, may provide some of the strongest measurements of dark matter for HAWC.
The M31 galaxy is close to the Milky Way, and its dark matter structure is well-known.
Due to the large halo of dark matter surrounding all galaxies, the M31 dark matter limits are as strong as those from the Segue 1 dwarf galaxy, though it is thirty times further away.
Similarly, the Virgo cluster at a distance of 16.8\,Mpc is nearly one thousand times further away than the dwarf galaxies.
However, due to expected substructure in galaxies and clusters, M31 and the Virgo cluster are some of the strongest targets for dark matter detection with HAWC.

Observations of M31~\cite{Dugger:2010ys} and the Virgo cluster~\cite{Han:2012uw} using the Fermi-LAT have been shown to provide strong dark matter limits.
Additionally, observations have been made at TeV energies of other galaxy clusters, including the Fornax cluster observations of H.E.S.S~\cite{Abramowski:2012au}.
In Figure~\ref{HAWCM31Virgo}, we show the dark matter limits from the H.E.S.S. Fornax cluster observations~\cite{Abramowski:2012au} and the Fermi-LAT observations of the Virgo cluster~\cite{Han:2012uw}.
However, because the analyses in these papers employed different substructure boost factors than we consider here, we have scaled the boost factors to those expected from Ref.~\cite{Sanchez-Conde:2013yxa} for comparison.
The HAWC limits from the Virgo cluster are stronger than both the Fermi-LAT Virgo cluster limits and the H.E.S.S. Fornax cluster limits for dark matter masses above $\sim10$ TeV for the $b\bar{b}$, $t\bar{t}$, $\mu^+\mu^-$, and $W^+W^-$ channels and above $\sim2$ TeV for the $\tau^+\tau^-$ channel.
Moreover, the HAWC substructure-boosted limits from M31 will provide the strongest dark matter cross-section limits from any extragalactic source above a few TeV in dark matter mass.

As can be seen from Figure~\ref{HAWCGC}, the GC is another excellent source for HAWC studies of dark matter.
Though the GC is over $48\degree$ from the zenith, the large dark matter $J$-factors from the GC cause its projected dark matter annihilation limit to be an order of magnitude more constraining than those from dwarf galaxies.
One feature that the HAWC analysis shares with other instruments of similar point-spread function is a sensitivity which is not very dependent on the chosen dark matter halo profile of a source.
For the GC analysis, for instance, the combination of the large extent of the Einasto profile and the large peaked flux of the NFW profile with our optimal angular binning actually produces very similar dark matter limits from the two profiles.
Because of the uncertainties about the shape of the dark matter profile toward the GC, such robust limits are very useful.
Though the GC limits from HAWC are slightly weaker than those from the substructure-boosted M31 for lower masses, the GC limits from HAWC will be the strongest HAWC limits above a few tens of TeV.

Measurements from the GC have given some of the strongest limits on the dark matter cross-section~\cite{Cirelli:2009dv,Baxter:2011rc,Hooper:2011ti,Abramowski:2011hc,Abazajian:2011ak,Abazajian:2014fta}.
For TeV dark matter masses, the 112-hour observations of the GC from H.E.S.S. provide the strongest measured limits on the dark matter cross-section.
Because of the large HAWC zenith angle of the GC, the HAWC GC limits will not be as constraining as those from H.E.S.S.~\cite{Abramowski:2011hc,Abazajian:2011ak} except at the highest dark matter masses under consideration.
However, for leptonic channels above a few hundred TeV and hadronic and bosonic channels above $\sim1000\ {\rm TeV}$, the HAWC GC limits should be comparable to the H.E.S.S. limits as shown in Figure~\ref{HAWCGC}.
A southern-hemisphere version of HAWC, additionally, would improve the HAWC sensitivity to the GC by over an order of magnitude and constrain the GC dark matter cross-section as strongly as H.E.S.S. down to a few TeV.

While the HAWC GC dark matter sensitivity should only be greater than that from H.E.S.S. for 100-1000 TeV, the HAWC sensitivity from sources with lower astrophysical backgrounds, such as M31, should be the strongest observed above a few TeV.
The HAWC cross-section limits should be a factor a few hundred from the thermal cross-section for most channels.
However, for the $W^+W^-$ dark matter channel, which has natural Sommerfeld enhancement, the HAWC sensitivity is within an order of magnitude of thermal from 20\,TeV to 400\,TeV and reaches the thermal value between 4-5\,TeV.
However, because many dark matter candidates in this parameter space are not thermally produced, the thermal cross-section should only be taken as a benchmark model, as dark matter candidates may have cross-sections above that value.
\subsection{Further Analyses with HAWC}
Due to the large HAWC field of view, additional HAWC analysis of dwarf galaxies can be performed as well.
Though Segue 1 is one of the best {\it known} dwarf galaxies in which to look for dark matter signals, HAWC can also look for emission from dwarf spheroidals which are currently {\it unknown}.
Dwarf galaxies are extremely faint, and the best candidate dwarf galaxies are those with the lowest luminosities, which have the highest dark matter mass and the lowest luminous matter backgrounds.
Therefore, it is likely that the best candidate dwarf galaxy for our analysis has not yet been discovered.
However, with the wide field-of-view of HAWC, it can search for faint gamma-ray signals with hard spectra in locations with no known counterparts, which would be the expected dark matter annihilation signal from an unknown dwarf galaxy.
The mass of the Segue 2 dwarf galaxy, for instance, was only recently measured due to its low luminosity~\cite{Kirby:2013isa}.
Similar searches for dark matter from undiscovered dwarfs have been done using {\it Fermi}-LAT data~\cite{Nieto:2013xaa}.
Additionally, Ref.~\cite{Harding:2013qra} recently showed that such a subhalo could be responsible for the TeV cosmic-ray anisotropy observed with Milagro~\cite{Abdo:2008kr}, and if so, HAWC should be able to detect gamma rays from such a dark subhalo within one year of operations.

In addition, HAWC can observe all known dwarf galaxies within its field of view and do a joint likelihood analysis of all the spectra.
Such an analysis was performed by the {\it Fermi}-LAT, which showed that combined limits from their considered 10 dwarf galaxies are more than twice as constraining on the dark matter signal than their strongest individual dwarf galaxy~\cite{Ackermann:2011wa,GeringerSameth:2011iw}.

A joint likelihood analysis of multiple dwarf galaxies would involve determining the probability of a given dark matter profile for each dwarf galaxy and comparing the expected photon fluxes from the considered galaxies to the observed photon counts by HAWC.
By varying the dark matter cross-section, the maximal likelihood could be determined, giving either a measurement of the WIMP cross-section or an upper limit.
Similar to the dwarf galaxies, HAWC can study a joint-likelihood of several galaxy clusters, giving an even better limit than from one cluster alone.

HAWC can search the parameter space of possible dark matter explanations of astrophysical excesses.
The AMS-02 positron excess, for instance, could be explained by a dark matter candidate with mass of 1.5-3 TeV and cross-section of $\langle\sigma_{\rm{A}}v\rangle=(6-23)\times10^{-24}\rm\,cm^3\,s^{-1}$ which annihilates into two pairs of muons and/or pions~\cite{Cholis:2013psa}.
Such a hard, high-cross-section signal should produce an observable signal in M31, and the HAWC substructure-boosted M31 limits could be able to verify or reject such explanations of the AMS-02 positron excess within five year of running.
\section{Conclusions}
The HAWC observatory will provide strong sensitivity to high-mass WIMP dark matter from dwarf spheroidal galaxies, M31, the Virgo cluster, and the GC after five years of observations.
In the $b\bar{b}$ and $t\bar{t}$ channel, HAWC should observe cross-sections down to $\langle\sigma_{\rm{A}}v\rangle_0\approx10^{-23}\rm\,cm^3\,s^{-1}$.
For the $\tau^+\tau^-$ channel, HAWC GC observations should be able to study dark matter cross-sections of $\langle\sigma_{\rm{A}}v\rangle_0\approx5\times10^{-24}\rm\,cm^3\,s^{-1}$ above 10\,TeV in the GC and cross-sections down to $\langle\sigma_{\rm{A}}v\rangle_0\approx3\times10^{-24}\rm\,cm^3\,s^{-1}$ below 10\,TeV in the substructure-boosted M31.
For the $W^+W^-$ channel with conservative natural Sommerfeld enhancement, HAWC should be able to constrain the dark matter cross-section to within a factor of 3 of thermal above 20\,TeV and probe thermal-scale dark matter between 4-5\,TeV with the GC and substructure-boosted M31.
In five years, HAWC can also verify with the $\mu^+\mu^-$ channel whether the observed AMS-02 positron excess is due to dark matter annihilation.

The projected limits discussed here are based on the current understanding of the HAWC detector performance.
With the operation of HAWC should come better understanding of its gamma/hadron separation and improvements on the energy resolution and angular resolution of the observatory.
Better angular resolution would help reject background photons outside of the region-of-interest for sources without significant extension, and good energy resolution is necessary to compare the spectral shape of any signal to that expected from the dark matter.
Greater understanding of backgrounds in the gamma-ray sky should improve the sensitivity of HAWC to point-sources, as will improved background-removal techniques.
Joint likelihood analyses should also improve the ability of HAWC to detect the dark matter.

HAWC is an excellent detector for searching for annihilating high-mass dark matter.
Its all-sky field-of-view and near-continuous observation time enables the observation of many dark matter sources and the ability to look for dark matter annihilations from previously unknown locations on the sky.
With the operation of the HAWC observatory, we can probe dark matter at higher masses with better sensitivity than ever before.

%
\begin{acknowledgments}
We acknowledge the support from: US National Science Foundation (NSF); US Department of Energy Office of High-Energy Physics; The Laboratory Directed Research and Development (LDRD) program of Los Alamos National Laboratory; Consejo Nacional de Ciencia y Tecnolog\'{\i}a (CONACyT), M\'exico; Red de F\'{\i}sica de Altas Energ\'{\i}as, M\'exico; DGAPA-UNAM IN108713 IG100414-3, M\'exico; Luc-Binette Foundation UNAM Postdoctoral Fellowship; and the University of Wisconsin Alumni Research Foundation.
KNA is supported by NSF CAREER Grant No. PHY-11-59224.
\end{acknowledgments}
%
\bibliography{bibliography}

%

\end{document}